\documentclass[aps,prd,nofootinbib,showpacs,preprintnumbers,amsmath,amssymb]{revtex4-1}
\usepackage{epsfig}
\usepackage{graphics}
\usepackage{graphicx}
\usepackage{amssymb}
\usepackage{latexsym,slashed}
\usepackage{multirow, ulem}
\usepackage{graphicx,color}
\def\be{\begin{equation}}
\def\ee{\end{equation}}
\def\bea{\begin{eqnarray}}
\def\eea{\end{eqnarray}}
\def\bes{\begin{subequations}}
\def\ees{\end{subequations}}
\def\gsim{~\rlap{$>$}{\lower 1.0ex\hbox{$\sim$}}\;}
\def\pslash{\displaystyle{\not}p}
\def\kslash{\displaystyle{\not}k}

\def\lslash{\displaystyle{\not}\ell}

\def\kpslash{\displaystyle{\not}k'}
\def\lpslash{\displaystyle{\not}\ell'}

\newcommand{\cF}{{\cal F}}
\newcommand{\bu}{{\overline u}}
\newcommand{\bv}{{\overline v}}
\newcommand{\bF}{{\overline F}}

\begin{document}

\title{ Electron-muon colliders at high energies  to discover heavy sterile neutrinos}

\author{Gorazd Cveti\v{c}$^a$}\email{gorazd.cvetic@usm.cl}
\author{Claudio Dib$^a$}\email{ claudio.dib@usm.cl}
\author{ C.~S.~Kim$^{b,c}$}\email{ cskim@yonsei.ac.kr, corresponding author}
\author{Vishnudath K.~N.$^d$}\email{ vishnudath.kn@vit.ac.in}

\affiliation{
$^a$ Department of Physics, Universidad T{\'e}cnica Federico Santa Mar{\'\i}a,  Casilla 110-V, Valpara{\'\i}so, Chile\\ 
$^b$  Department of Physics and IPAP, Yonsei University, Seoul 03722, Korea \\
$^c$  CPNR, Department of Physics, Chonnam National University, Gwangju 61186, Korea \\
$^d$ Department of Physics, School of Advanced Sciences, Vellore Institute of Technology,
Vellore 632014, Tamilnadu, India}

\begin{abstract}

We study high-energy charged-lepton-flavor-violating (cLFV)  channels in $e^- \mu^+$ scattering to discover heavy sterile neutrinos, which appear naturally as a minimal extension of the Standard Model.  For $\sqrt{s} \le 2M_W$, we consider the process $e^- \mu^+ \to e^+ \mu^-$, which is dominated by one-loop box diagrams. We numerically evaluate the box diagrams, which involve a high-energy extension of the Inami–Lim functions, and find that the amplitudes are strongly suppressed due to their quartic dependence on light–heavy mixing.  For numerical evaluations we obtain the maximal possible rates given the current active light to heavy sterile neutrino mixing bounds. For $\sqrt{s} > 2M_W$, we analyze $e^- \mu^+ \to W^+ W^-$ and compute cross sections in both single-sterile and minimal type-I seesaw scenarios. We find this latter process to be more promising for the evidence of heavy sterile neutrinos in $e$--$\mu$ colliders.

\end{abstract}
\maketitle

\label{sec:Intro}

%
%

---

\section{Introduction}


Muon-based accelerator facilities have recently attracted renewed attention as promising tools for precision measurements and discovery physics \cite{Delahaye:2019MuonColliders, Bartosik:2020MuonColliderStudy, Blondel:2020MuonColliderFrontiers, MuonColliderCollaboration:2022ForumReport}. A high-energy muon collider combines several advantages over $e^+e^-$ machines: reduced synchrotron radiation, the possibility of reaching multi-TeV center-of-mass energies within a more compact ring, and a clean leptonic initial state that allows precise theoretical predictions. In addition to muon-antimuon collisions, asymmetric lepton beams, including electron–muon colliders, have been proposed \cite{Choi:1997bm, Bossi:2020yne, Hamada:2022mua, Lu:2020dkx}. An electron–muon collider would offer a qualitatively new probe of electroweak interactions by directly accessing scattering processes between distinct charged leptons. Such a machine naturally provides sensitivity to flavor-dependent new physics and enables direct tests of { charged-lepton-flavor-violation (cLFV) } in high-energy processes. Unlike low-energy searches, which probe cLFV through rare decays, high-energy electron–muon scattering allows us to study the origin of flavor violation directly through production cross sections, angular distributions, and energy scaling behavior.
{An electron-muon collider would thus provide a unique experimental environment, probing interactions between distinct lepton flavors at high energies and offering direct sensitivity to flavor-violating phenomena \cite{Choi:1997bm}.}

Charged lepton flavor violation (cLFV)  \cite{Kuno:1999, Raidal:2008jk, Calibbi:2017uvl, Lindner:2016bgg, Antusch:2014woa, delAguila:2008cj, Deppisch:2015qwa} constitutes one of the most promising avenues for discovering new physics beyond the Standard Model of Particle Physics (SM).
%
%
Although the SM has achieved remarkable success, providing 
a precise description of fundamental particles and their interactions, and
withstanding 
stringent experimental tests, it is widely recognized that it must be incomplete. In particular, it offers no viable candidate for dark matter \cite{Bertone:2004pz, Feng:2010gw, Arcadi:2017kky, Roszkowski:2017nbc, Alexander:2016aln} and fails to account for the observed baryon asymmetry of the universe \cite{Riotto:1999yt, Dine:2003ax, Buchmuller:2005eh, Davidson:2008bu, Canetti:2012zc}. These shortcomings strongly motivate the search for physics beyond the SM. In this respect, the first clear experimental indication of {how to modify the SM came} 
through the observation of neutrino oscillations \cite{Fukuda:1998mi, Ahmad:2001an, Ahmad:2002jz, Eguchi:2002dm, Ahn:2006zza, Adamson:2008zt, Abe:2011sj, Abe:2011fz,  Ahn:2012nd, An:2012eh, Adamson:2013whj}, which imply that neutrinos are massive and that 
 lepton flavors should mix —phenomena that cannot be accommodated within the minimal SM framework. Within the SM augmented solely by light neutrino masses, cLFV processes are suppressed to unobservable levels \cite{Petcov:1976ff, Bilenky:1977du, Lee:1977tib}. Consequently, any observable cLFV signal at colliders would constitute an indication of new degrees of freedom. From a phenomenological perspective, collider observables sensitive to cLFV include flavor-changing scattering channels, deviations in electroweak cross sections induced by non-unitary lepton mixing, and the resonant or non-resonant exchange of new neutral leptons. Heavy neutral leptons (HNLs) 
appear naturally in seesaw mechanisms introduced to explain the origin of neutrino masses \cite{Minkowski:1977sc,Yanagida:1979as,GellMann:1979vob,Mohapatra:1979ia}. In type-I seesaw models and their low-scale variants 
\cite{Mohapatra:1986bd,Wyler:1982dd,Bernabeu:1987gr,Akhmedov:1995vm}, gauge-singlet { (also called {\sl sterile})} fermions mix with the active neutrinos, generating small neutrino masses while allowing sizable active–sterile mixing angles compatible with current experimental bounds. If HNLs have masses  near the electroweak or TeV scale, they can contribute directly to high-energy scattering amplitudes through both on-shell production and virtual exchange, and induce cLFV as well.

 Electron–muon collisions are especially sensitive to heavy sterile neutrinos. The flavor structure of the initial state probes directly the mixing parameters connecting the electron and muon sectors, which are otherwise constrained mainly through rare decay experiments. Heavy sterile neutrinos may mediate cLFV reactions such as
\[
e^- \mu^+ \rightarrow W^- W^+\quad \text{,}\quad 
e^- \mu^+ \rightarrow \ell_i^\pm W^\mp \quad \text{{or}}\quad
e^- \mu^+ \rightarrow e^+ \mu^-  ,
\]
or induce measurable distortions in Standard Model channels via modified charged-current interactions. Depending on the sterile neutrino mass relative to the collider energy, one expects qualitatively different regimes: resonant production when kinematically accessible, threshold enhancements near production onset, and contact-interaction–like behavior for heavy states well above the collision energy.

A distinctive feature of heavy sterile neutrino scenarios is the interplay between high-energy collider observables and neutrino mass generation mechanisms. The same parameters controlling light-neutrino masses also determine the active–sterile mixing, cLFV rates, and deviations from unitarity of the leptonic mixing matrix. Electron–muon colliders therefore provide a complementary probe to intensity-frontier experiments, allowing simultaneous tests of the flavor structure and mass scale underlying the seesaw framework.

In this work, we investigate $e^-\mu^+$ collisions at high energies within the framework of heavy sterile neutrino models, to assess the discovery potential of future $e^-\mu^+$ colliders. In particular we study signatures of {charged} lepton flavor violation and deviations from Standard Model expectations. We first study the process $e^-\mu^+\to e^+\mu^-$ at energies below $2M_W$.   Having no neutrinos in the final state, this process violates lepton flavor. However, the SM process $e^-\mu^+\to e^+\mu^- \nu_e \bar\nu_\mu$ which is certainly dominant and conserves lepton flavor, is a background for the neutrinoless $e^-\mu^+\to e^+\mu^-$, because the neutrino pair in the final state goes undetected, except for the missing energy. Our cLFV neutrinoless process would only show as an extra contribution at the end of the $e^+\mu^-$ invariant mass spectrum in $e^-\mu^+\to e^+\mu^- \nu_e \bar\nu_\mu$. In scenarios with HNLs, this neutrinoless process appears at 1 loop order (``box'' diagrams). For higher energies, namely $\sqrt{s} > 2 M_W$, we study the process $e^-\mu^+ \to W^+W^-$, where the on-shell $W$'s decay into their standard modes, i.e pure leptonic ($\ell^+\nu_\ell + \ell^{\prime -}\bar\nu_{\ell '}$), pure hadronic ($u_i \bar d_j + \bar u_k d_l$) or mixed ($\ell^+\nu_\ell + \bar u_i d_j$ or $\ell^-\bar\nu_\ell + u_i \bar d_j$). While all these processes violate lepton flavor, in the leptonic modes the neutrinos escape detection, so that the $W$ masses cannot be easily reconstructed.

This paper is organised as follows. In Sec.~\ref{sec:Box} we calculate the cLFV process $e^- \mu^+ \to e^+ \mu^-$, i.e., the purely leptonic process where there is neither missing energy (light neutrinos) nor hadrons in the final state. In the calculation, a (numerical) extension of the Inami-Lim functions is needed, because the center-of-momentum (CM) energies considered here are large, namely $m_e \ll \sqrt{s}$ ($\leq 2 M_W$). In Sec.~III we consider the scattering of $e^- \mu^+$ at larger CM energies, $\sqrt{s} > 2 M_W$, where the channel $e^- \mu^+ \to W^+ W^-$ opens up and becomes dominant, and the $W$'s promptly decay either leptonically or hadronically. In Sec.~IV we summarise our results. Appendix A contains some details on the extended Inami-Lim functions and Fierz transformations for the box diagrams.


\section{Process $e^- \mu^+ \to \mu^- e^+$ for $\sqrt{s} < 2 M_W$  via box diagrams}
\label{sec:Box}

\begin{figure}[h]
\begin{center}
\includegraphics[width=0.3\textwidth]{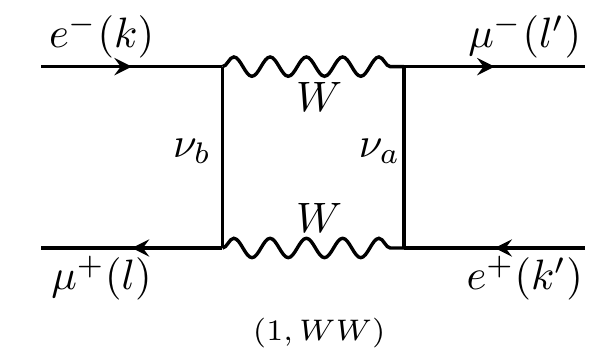} ~~
\includegraphics[width=0.3\textwidth]{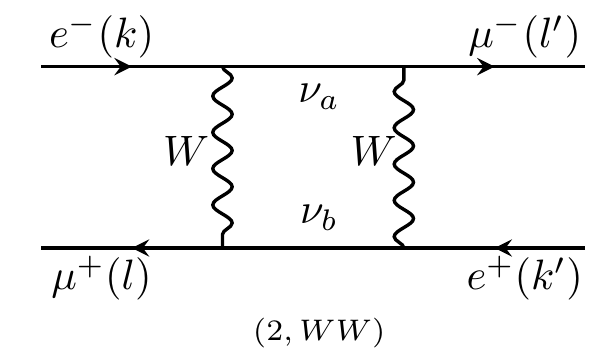} \\
\vspace{0.8cm}
\includegraphics[width=0.3\textwidth]{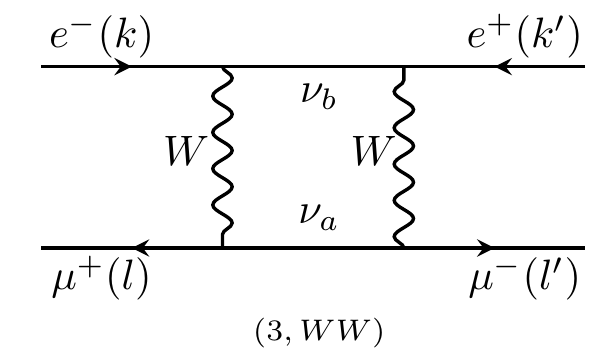} ~~
\includegraphics[width=0.3\textwidth]{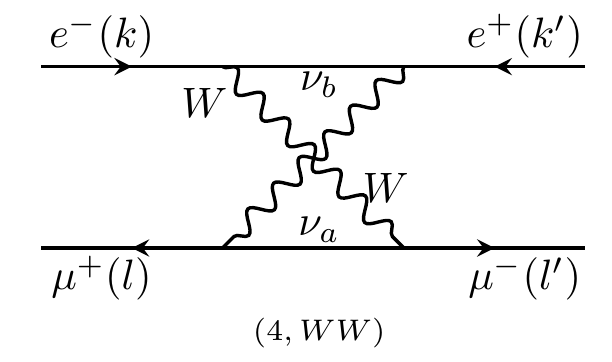}
\end{center}
\caption{\footnotesize The box diagrams for the scattering process $e^-(k) \mu^+(\ell) \to \mu^-(\ell') e^+(k')$, {of the types $(t, WW)$ ($t=1,2,3,4$), where $WW$ indicates that these are the box diagrams with two $W$-propagators.} The boxes also contain two neutrino propagators (of neutrinos $\nu_a$ and $\nu_b$ that can be light or heavy). {The diagrams of the type $(t,p)$ with $p=GW, WG, GG$ (and same types $t=1,2,3,4$) are those where one or both $W$-propagators are replaced by the Goldstone propagator (see the text).}}
\label{FigBox}
\end{figure}

First we discuss the scattering of $e^-$ and $\mu^+$ at high energies $\sqrt{s} \gg m_{\mu}$, but under the threshold value $\sqrt{s} < 2 M_W$. In a minimally extended SM, i.e., SM with three light and a number $n_H$ of heavy {sterile} neutrinos, the process $e^-(k) \mu^+(\ell) \to \mu^-(\ell') e^+(k')$ appears at one-loop, with the four box diagrams depicted in Fig.~\ref{FigBox}. The 
{one-loop} penguin diagrams 
{do not} contribute to that process.
\footnote{See the comments at the end of this Section.}

We work in the Feynman gauge ($\xi=1$), and 
we consider first that the two { neutrinos } $\nu_a$ and $\nu_b$, whose propagators appear in the box diagram, have the usual SM coupling with $W$ and the charged leptons $e$ and $\mu$, disregarding the mixing elements.
The effects of the heavy-light neutrino mixings will be incorporated later in a straightforward way. 
{In the Feynman gauge, in addition to the box diagrams depicted in Fig.~\ref{FigBox}, are the diagrams where one or both $W$'s are replaced by the corresponding charged would-be Goldstone boson. 
This means that in the Feynman gauge we do not have four but 16 diagrams altogether.
}


{The corresponding amplitude will be denoted as $A_{\rm box}$,
which
is the sum of the 16 box diagram contributions mentioned above, ${\cal A}_{\rm box}=\sum {\cF}^{(t, p)}_f$,  where the index $t=1,2,3,4$ refers to any of the four types of diagrams 
(``topologies'') depicted in Fig.~\ref{FigBox},  while the index $p= WW, GW, WG, GG$ refers to the internal bosons in the box ($W$ or Goldstones), and the subscript ``$f$'' indicates that the heavy-light neutrino mixings are included. }
%
%
{We will first consider no heavy-light neutrino mixing effects --thus no subscript ``$f$''--, incorporating them later. In our convention, the reduced amplitude 
$\cF^{(1,{WW})}$ corresponds to the first diagram in Fig.~\ref{FigBox}, i.e., with two $W$ propagators. $\cF^{(1,{GW})}$ and $\cF^{(1, WG)}$ are the amplitudes corresponding to the analogous diagrams where one of the $W$-propagators is replaced by the Goldstone propagator, and $\cF^{(1,GG)}$ corresponds to the analogous diagram where both $W$-propagators are replaced by the Goldstone propagators. Analogous conventions are adopted for the reduced amplitudes $\cF^{(t,p)}$ of topologies $t=2, 3, 4$.
}

Accordingly, the expression for the reduced amplitude 
$\cF^{(1,{WW})}$ 
is the following:
\bea
\lefteqn{\cF^{(1,{WW}) = \left( \frac{g}{\sqrt{2}}\right)^4 \times  }  }
\nonumber\\ &&
\int \frac{d^4 p}{(2 \pi)^4} \frac{(-1) \left[ \bu(\ell') \gamma^{\alpha} P_L (\pslash +\lpslash +M_a) \gamma^{\beta} P_L v(k') \right] \left[ \bv(\ell) \gamma_{\beta} P_L (\pslash + \kslash + M_b) \gamma_{\alpha} P_L u(k) \right]}{ \left((p+\ell')^2-M_a^2 +i \Gamma_a M_a \right) \left( (p+k)^2 - M_b^2 +i \Gamma_b M_b \right) \left( (p+k+\ell)^2 - M_W^2 +i \Gamma_W M_W \right) \left( p^2 - M_W^2 +i \Gamma_W M_W \right) },
\nonumber\\
\label{F1a1} \eea
where $M_a$ and $M_b$ are the masses of the two neutrinos (and $\Gamma_a$ and $\Gamma_b$ are their corresponding decay widths). The spinor normalisation conventions used are
\be
\sum_{h} u(k) \bu(k) = (\kslash+m_e); \quad {\rm etc.},
\label{normu} \ee
and $P_L=(1-\gamma_5)/2$, $P_R=(1+\gamma_5)/2$ are the two (left-hand and right-hand) 
chirality projectors.
In the process, the Mandelstam variables are
\be
(k+ \ell)^2 = s; \; (k-k')^2=t; \; (k- \ell')^2= u. 
\label{Mand} \ee
We will assume that $s \gg m^2_{\mu}$ ($\sim 10^{-2} \ {\rm GeV}^2$), i.e., we will consider everywhere $m_e, m_{\mu} \mapsto 0$. As a consequence, we have $s+t+u=0$.
The use of the Feynman-Schwinger parameters, and algebraic manipulations of the numerator expression in the integrand of Eq.~(\ref{F1a1}), including Fierz transformations, then lead to the following expression for the above amplitude: 
\bea
\lefteqn{
\cF^{(1,WW)} =
}  
\nonumber\\ &&
\frac{i}{16 \pi^2} \frac{g^4}{2 M_W^2} \left\{
\bF^{(1)}_{1} \left[ \bu(\ell') \gamma^{\mu} P_L v(k') \right] \left[ \bv(\ell) \gamma_{\mu} P_R u(k) \right] - 2 \bF^{(1)}_2(1-x,1-y) \frac{1}{M_W^2} \left[\bu(\ell') \kslash P_L v(k') \right] \left[ \bv(\ell) \lpslash P_L u(k) \right] \right\},
\nonumber\\
\label{F1a2}
\eea
where $\bF^{(1)}_{j}$ ($j=1,2$) are integrals over the Feynman-Schwinger parameters related to the box diagrams of type $t=1$:


 \be
\bF^{(1)}_1 = \int dx\, dy\, dz \frac{1}{\Delta^{(1)}_W}, \quad
\bF^{(1)}_2(f_1,f_2) =    \int dx\, dy\, dz \frac{f_1 f_2}{\left( \Delta^{(1)}_W \right)^2},
\label{bF1def} \ee
where the integration is over $0 \leq x + y + z \leq 1$ and $0 \leq x,y,z \leq 1$, and the expression $\Delta^{(1)}_W$ in the denominators is
\be
\Delta^{(1)}_W =\left[ (1-x-y) (1 - s_W z) + s_W z^2 + x y (- u_W) + x x_a + y x_b \right] - i \varepsilon_W - i \varepsilon_N.
\label{DeltaW1} \ee
Here, we denoted
\bes
\label{denot}
\bea
s_W & = & \frac{s}{M_W^2}, \; u_W = \frac{u}{M_W^2}, \;  t_W = \frac{t}{M_W^2},
\label{sWuWtW} \\
x_a & = & \frac{M_a^2}{M_W^2}, \; x_b = \frac{M_b^2}{M_W^2},   \;  x_W \equiv M_N^2/M_W^2,
\label{xaxb} \\
\varepsilon_W & = & \frac{\Gamma_W}{M_W} (1 - x - y), \;
\varepsilon_N = x \frac{\Gamma_a M_a}{M_W^2} + y \frac{\Gamma_b M_b}{M_W^2}.
\label{eps} \eea \ees
We note that, since in our approximation, $m_e, m_{\mu} \mapsto 0$, we have $k^2=\ell^2=k'^2 = \ell'^2 =0$, $s= 2 k \cdot \ell = 2 k' \cdot \ell'$, etc.
In the center of mass frame (CM), if we denote by $\theta$ the angle of the outgoing $\mu^-$ with respect to the 
incoming $e^-$, then we have
\be
u_W = - \frac{1}{2} s_W (1 - \cos \theta), \;
t_W = - \frac{1}{2} s_W (1 + \cos \theta).
\label{uWtW} \ee



Similar expressions are obtained for the diagrams of the types ($t$,{\small WW}), ($t$,{\small WG}), ($t$,{\small GW}) and ($t$,{\small GG}), where $t=1,2,3,4$ according to Fig.~\ref{FigBox}. The expressions for all these diagrams
are given in Appendix \ref{app:Box}, where we also included the Fierz transformations that help transform all the terms quartic in the spinors (in the amplitudes) to expressions  made of the spinors $u$ (not $v$): $u(k)$, $u(\ell')$, $\bu(\ell')$ and $\bu(k')$. Such a structure enables then a direct evaluation of the traces appearing in the squared amplitude $\langle |\cF|^2 \rangle$.

Let us now incorporate the heavy-light mixing parameters in the above amplitudes.
In this scenario of heavy-light mixing, with $n_H$ heavy neutrino states, the flavor neutrino eigenstates $\nu_{\eta}$ ($\eta=e, \mu, \tau$) are related to the neutrino mass eigenstates $\nu_a$ ($a=1,2,3,\ldots,3+n_H$) by the mixing maxtrix $U$ as
\begin{equation}
\nu_{\eta} = \sum_{a=1}^{3+n_H} U_{\eta a}\, \nu_a = \sum_{a=1}^{3} U_{\eta a}\, \nu_a +
\sum_{c=1}^{n_H} U_{\eta (3+c)}\, N_c,
\label{hlmix} 
\end{equation}
where $\nu_{3+c} = N_c$ ($c=1,\ldots,n_H$) are the heavy mass eigenstates. In our numerical evaluations, we will work in the simplified scenario of only one massive sterile neutrino $N_c=N$ (i.e., $n_H=1, U_{\eta 4} \equiv U_{\eta N}$). Therefore, the neutrino masses are taken as $M_a =0$ 
for $a=1,2,3$, and 
$M_N$ for $a=4$ (i.e., $c=1$). Further, for simplicity, we will ignore throughout the CP-violating phases, as well as the Majorana creation phase factors ($\lambda_j \mapsto 1$).

 The mixing implies that each amplitude $\cF^{(t,p)}(M_a,M_b)$ (here $a,b = 1,2,3,4$) is multiplied by:  (i) the factor $U_{e a}^{\ast} U_{\mu a} U_{e b}^{\ast} U_{\mu b}$ for the diagrams of type $t=1,2$, and (ii) the factor  $U_{\mu a} U_{\mu a} U_{e b}^{\ast} U_{e b}^{\ast}$ for types $t=3,4$.
 Notice that when any of the neutrinos $\nu_a$ or $\nu_b$ is massless, the only amplitudes that do not vanish are $\cF^{(t,WW)}(M_a,M_b)$ for $t=1,2$. Indeed, the diagrams of type $t=3,4$ are only for Majorana neutrinos, and all diagrams that contain a Goldstone boson vanish in such a case because their coupling being proportional to the fermion mass. 
 


We consider the three light neutrinos $\nu_a$ as massless (for all practical purposes): $M_a=0$ ($a=1,2,3$). Also, due to the unitarity of the neutrino mixing matrix, we have
 \be
 \sum_{a=1}^3 U_{e a}^{\ast} U_{\mu a} = - U_{e N}^{\ast} U_{\mu N}.
 \label{Bunit} \ee
 This then implies that the amplitudes $\cF^{(t,p)}(M_a,M_b)$ with $t=1,2$ appear in the following combination once the mixing is included (thus the subscript ``$f$'' from {\sl full} amplitudes $\cF^{(t,p)}_f$):
 \bea
 \cF^{(t,p)}_f & = & \sum_{a=1}^4 \sum_{b=1}^4 
 U_{e a}^{\ast}  U_{\mu a} U_{e b}^{\ast} U_{\mu b} \ 
 \cF^{(t,p)}(M_a,M_b) 
 \nonumber \\
 & = & (U_{e N}^{\ast} U_{\mu N})^2 \left[ \cF^{(t,p)}(0,0) - \cF^{(t,p)}(M_N,0) - \cF^{(t,p)}(0, M_N) + \cF^{(t,p)}(M_N,M_N) \right] \quad (t=1,2).
 \label{cFj12} \eea
 The amplitudes $\cF^{(t,p)}(M_a,M_b)$ for $t=3,4$ are nonzero only when $M_a=M_b=M_N$, and therefore
  \bea
  \cF^{(t,p)}_f & = & \sum_{a=1}^4 \sum_{b=1}^4 
  (U_{\mu a})^2 (U_{e b}^{\ast})^2 \ 
  \cF^{(t,p)}(M_a,M_b) 
  \nonumber \\
  & = &  (U_{e N}^{\ast} U_{\mu N})^2\  \cF^{(t,p)}(M_N,M_N) \quad (t=3,4).
  \label{cFj34} \eea

 The total cross section is then
 \be
 \sigma(e^- \mu^+ \to \mu^- e^+) = \frac{1}{32 \pi s} 
 \int_{-1}^{+1} d (\cos \theta) \left\langle {\bigg |} \sum_{t,p} \cF^{(t,p)}_f {\bigg |}^2 \right\rangle,
 \label{sigma} 
 \ee
 where $\theta$ is the angle  between the outgoing $\mu^{-}(\ell')$ and incoming $e^{-}(k)$ in the CM, cf.~Eq.~(\ref{uWtW}). The (many) traces appearing in $\langle |\sum \cF^{(t,p)}|^2 \rangle$ were performed using FeynCalc \cite{Mertig:1991FeynCalc,Shtabovenko:2016FeynCalc,Shtabovenko:2020FeynCalc,Shtabovenko:2023FeynCalc}.

  The value of the mixing parameter appearing at the box amplitude was taken to be
 \be
 | U_{\mu N} U_{e N}^{\ast} |^2 = 10^{-10},
 \label{mixnum} \ee
 which is consistent with the present upper bounds on $|U_{\mu N}|^2$ and $|U_{e N}|^2$, as shown in Fig.~\ref{fig:nonunitary}.

 The evaluation of the integrals over the Feynman-Schwinger parameters, Eqs.~(\ref{bFjdef}), was performed by Monte-Carlo (MC) evaluation using Mathematica \cite{Mathematica}. The value of $\varepsilon_N$, Eq.~(\ref{eps}), appearing in the denominators of these integrals, 
 in the present scenario of one sterile neutrino is reduced to
 \be
 \varepsilon_N = (x+y) \frac{\Gamma_N M_N}{M_W^2},
 \label{eps1N} \ee
 and was taken in our specific calculations to be $\varepsilon_N = 0.005$. We checked that MC evaluations do not change significantly when $\varepsilon_N$ is varied about that value. If we decreased $\varepsilon_N$ much further from this value, MC calculations often became unstable.

\begin{figure}[htb] 
\centering\includegraphics[width=100mm]{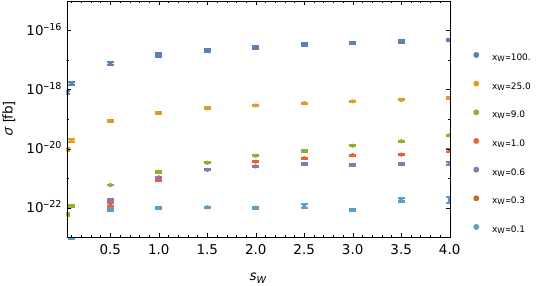}
\caption{\footnotesize The total cross section $\sigma(e^- \mu^+ \to \mu^- e^+)$, in pb, as a function of $s_W=s/M_W^2$, for various values of the heavy sterile neutrino mass $M_N$ ($x_W \equiv M_N^2/M_W^2$). The heavy-light mixing parameter combination was taken to be $| U_{\mu N} U_{e N}^{\ast} |^2 = 10^{-10}$.}
\label{FigsigBox}
 \end{figure}

 In Fig.~\ref{FigsigBox} we present the results for the cross section $\sigma(e^- \mu^+ \to \mu^- e^+)$ for $0 < s < 4 M_W^2$, and various values of the heavy neutrino masses $M_N$.
 We see that the cross section increases when the mass of the sterile neutrino increases, and when the CM energy $s$ increases. However, the values are highly suppressed by the heavy-light mixings, because $\sigma(e^- \mu^+ \to \mu^- e^+) \propto | U_{\mu N} U_{e N}^{\ast} |^4$. For a value $| U_{\mu N} U_{e N}^{\ast} |^2 = 10^{-10}$, which is approximately at the allowed upper bound, we obtain for $M_N = 10 M_W$ 
 ($x_W \equiv M_N^2/M_W^2 =100.$) and at $s = 4 M_W^2$ the values  $\sigma(e^- \mu^+ \to \mu^- e^+) \sim 10^{-16} \ {\rm fb}$, which is probably too low for the foreseeable future $e$-$\mu$ colliders.

{In contrast,} for $s_W=s/M_W^2 > 4$ the channel of $W^+W^-$ production on shell opens up ($e^- \mu^+ \to W^+ W^-$) which is a tree-level process mediated by neutrino exchange, for which the heavy-light neutrino suppression effects are considerably less severe, $\sigma(e^- \mu^+ \to W^+ W^-) \propto | U_{\mu N} U_{e N}^{\ast} |^2$. 
We will address this regime in the next section.

 In our calculations, we also checked numerically that for $s \ll M_W^2$ , i.e. when $s \to 0$ (and thus $u,t \to 0$),
 the box-loop integrals shown in Eq.~(\ref{bFjdef}) reduce to the known Inami-Lim functions. For example, we get
 \bes
 \label{bFs0}
 \bea
 \bF^{(t)}_1 {\Big |}_{s \to 0} \equiv \int dx\, dy\,  dz \frac{1}{\Delta^{(t)}_W} {\Bigg |}_{s \to 0}
 & \approx & (1/2) F_2(x_a,x_b),
 \label{bF1s0} \\
  \bF^{(t)}_2(1,1){\Big |}_{s \to 0} \equiv \int dx \, dy\,  dz \frac{1}{\left( \Delta^{(t)}_W \right)^2}{\Bigg |}_{s \to 0}  & \approx & - F_0(x_a,x_b),
  \label{bF2s0} \eea \ees
 where $F_2$ and $F_0$ are known Inami-Lim functions \cite{Inami:1980fz}
 \bes
 \label{IL}
 \bea
 F_2(x_a,x_b) & \equiv & i 16 \pi^2 M_W^2 \int \frac{d^4  p}{(2 \pi)^4} \frac{ p^2}{(p^2 - M_a^2) (p^2 - M_b^2) (p^2 - M_W^2)^2}
 \nonumber\\
 & = & \frac{1}{(x_a-x_b)} \left[ \frac{1}{(1-x_a)} + \frac{x_a^2}{(1-x_a^2)} \ln x_a - (x_a \leftrightarrow x_b) \right]
 \label{ILF2} \\
 F_0(x_a,x_b) & \equiv & i 16 \pi^2 M_W^2 \int \frac{d^4  p}{(2 \pi)^4} \frac{1}{(p^2 - M_a^2) (p^2 - M_b^2) (p^2 - M_W^2)^2}
 \nonumber\\
 & = & \frac{1}{(x_a-x_b)} \left[ \frac{1}{(1-x_a)} + \frac{x_a}{(1-x_a^2)} \ln x_a - (x_a \leftrightarrow x_b) \right]
 \label{ILF0} .
 \eea \ees
{
In our numerical evaluations we also found, as it should, that the approximate relations in Eqs.~(\ref{bFs0}) for $s\to 0$, become even more accurate if we also set the width parameters 
$\varepsilon_W, \varepsilon_N \to 0$ [cf.Eqs.~(\ref{eps})].
}
{ For our evaluation of the process $e^- \mu^+ \to \mu^- e^+$  we have derived and used the generalised Inami-Lim functions, Eqs. (\ref{bF1def}) and (\ref{bFjdef}). }

\begin{figure}[htb] 
\centering\includegraphics[width=90mm]{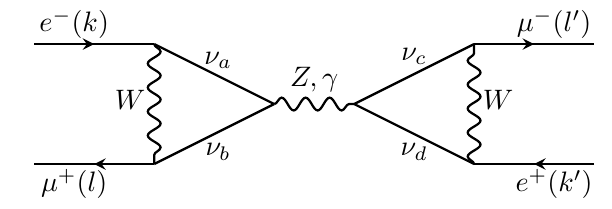}
\caption{\footnotesize A penguin diagram that contributes to the process $e^- \mu^+ \to mu^- e^+$.}
\label{FigPeng}
 \end{figure}

Finally, in Fig.~\ref{FigPeng} we show a typical ``double'' penguin diagram that contributes to the process $e^- \mu^+ \to \mu^- e^+$ we are considering. As shown, this is a two-loop  amplitude and therefore it is suppressed in comparison with the box diagrams (which are one-loop) by $\sim g^2$.
 Furthermore, it can be checked that the mixing factors in these amplitudes are again quartic in the heavy-light mixings, like in Eqs.~(\ref{cFj12})-(\ref{cFj34}).\footnote{See similar types of mixings that appear in the (one-loop) penguin diagrams for the process $e^- \mu^+ \to e^+ e^-$, \cite{Cvetic:2006yg}.} Therefore, in the scattering $e^- \mu^+ \to \mu^- e^+$ we did not consider the double-penguin diagram contributions.

{Now, if instead of the final state $\mu^- e^+$, we consider $e^- e^+$  or $\mu^- \mu^+$, the usual penguin-type diagram, which has just one loop, will be an important contribution, 
and the corresponding amplitude would be only quadratic in the heavy-light mixing parameters $U_{\ell N}$. Therefore, we expect that the cross section of the process $e^- \mu^+ \to e^- e^+$, at the considered energies $m_{\mu} \ll \sqrt{s} \leq 2 M_W$, would be much larger than 
$e^- \mu^+ \to \mu^- e^+$. The evaluation of this process is also not straightforward, because the explicit formulas for such penguin diagrams are available only in the zero energy limit \cite{Inami:1980fz}. It is interesting, however, that for $e^- \mu^+ \to e^- e^+$ the box diagrams also give some contributions that are quadratic (and not just quartic) in the mixing parameters $U_{\ell N}$. Namely, the full amplitudes $\cF^{(t,p)}_f$ with $t=1,2$ are obtained for such process by replacing $U_{\mu b} \mapsto U_{e b}$ in the first line of Eq.~(\ref{cFj12}), and then using the unitarity of the $U$ matrix; we thus obtain
 \bea
 \cF^{(t,p)}_f(e \mu \to e e) & = &
 (U_{e N}^{\ast} U_{\mu N}) |U_{e N}|^2 \left[ \cF^{(t,p)}(0,0) - \cF^{(t,p)}(M_N,0) - \cF^{(t,p)}(0, M_N) + \cF^{(t,p)}(M_N,M_N) \right]
 \nonumber\\ &&
 + (U_{e N}^{\ast} U_{\mu N}) \left[-\cF^{(t,p)}(0,0)+\cF^{(t,p)}(M_N,0) \right]
 \quad (t=1,2).
 \label{cFj12ee} \eea
The expression for the $t=3,4$ diagrams remains 
as in Eq.~(\ref{cFj34}), with $(U_{e N}^{\ast} U_{\mu N})^2$ replaced by $(U_{e N}^{\ast} U_{\mu N})  |U_{e N}|^2$.}

\section{Process $ e^- \mu^+ \rightarrow W^- W^+$ for $\sqrt{s} > 2 M_W$}\label{sec:WW}

In this section we consider the $e^-\mu^+$ collisions when the total CM energy $\sqrt{s}$ is greater than $2M_W$, resulting  in the production of on-shell $W^+ W^-$. 
{At tree level, this process can be mediated only by 
the active light neutrinos or by sterile neutrinos that have nonzero mixing with the 
active neutrinos.}

\begin{figure}[h]
\begin{center}
\includegraphics[width=0.2\textwidth]{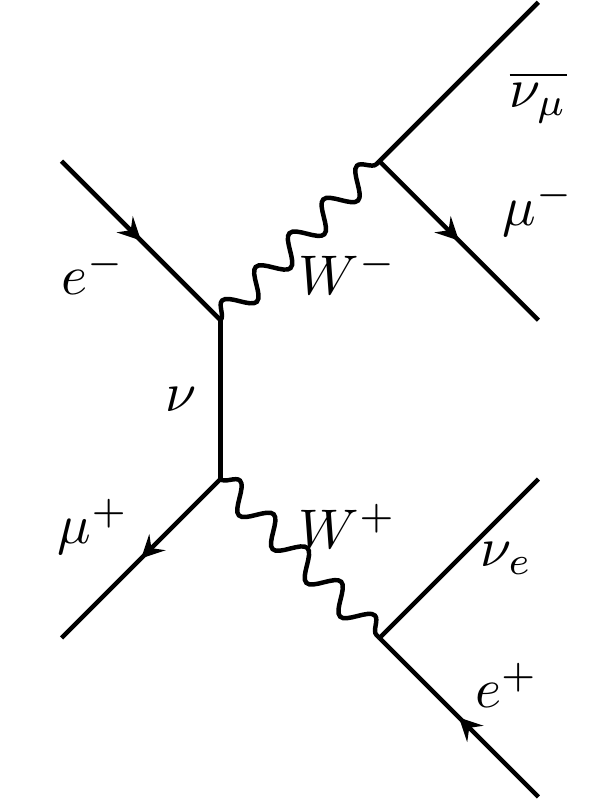} ~~
\includegraphics[width=0.2\textwidth]{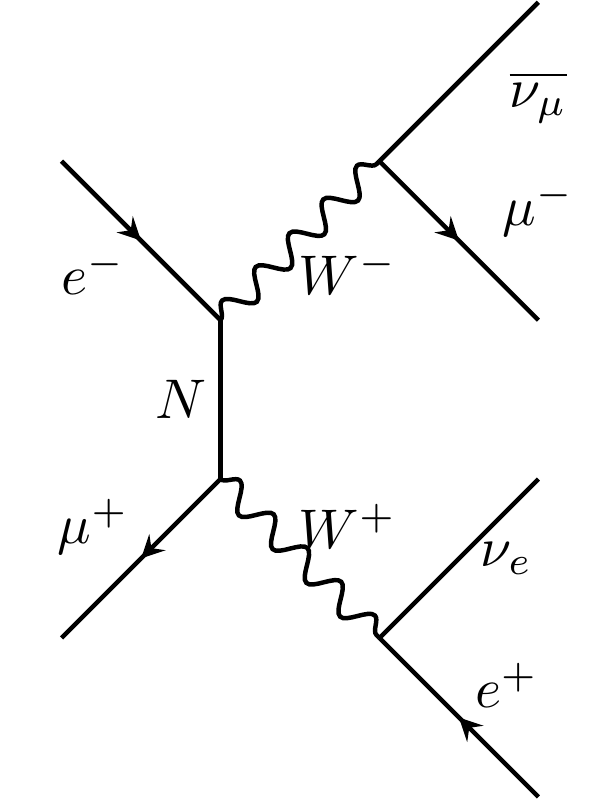} ~~
\includegraphics[width=0.2\textwidth]{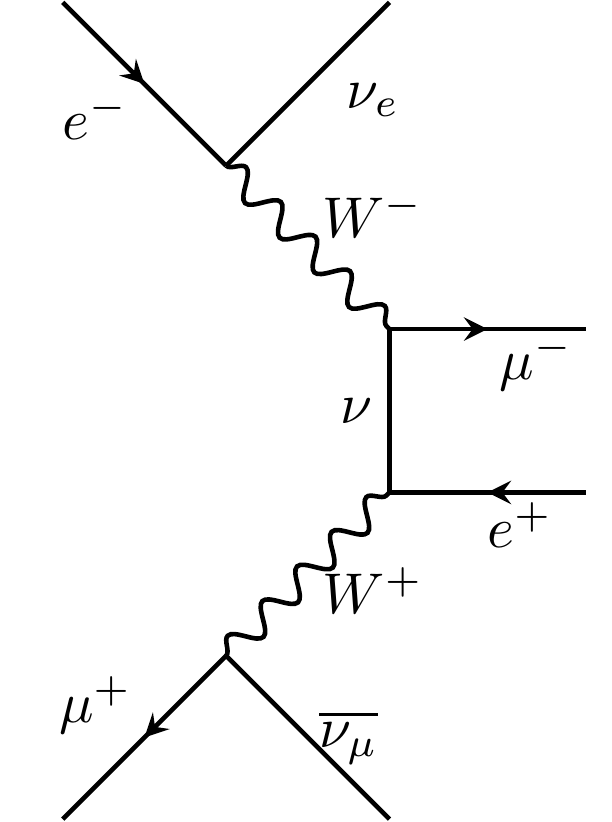} ~~
\includegraphics[width=0.2\textwidth]{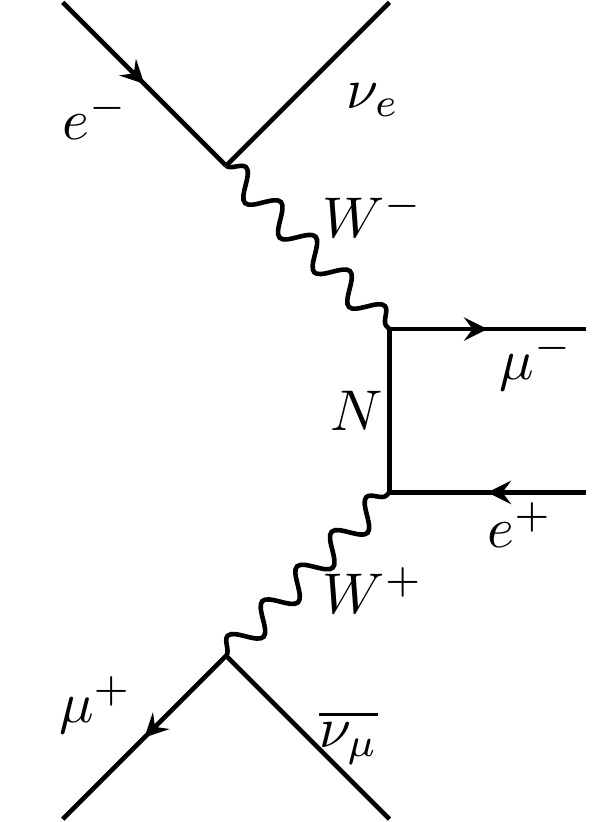} \\
\vspace{0.5cm}
\includegraphics[width=0.2\textwidth]{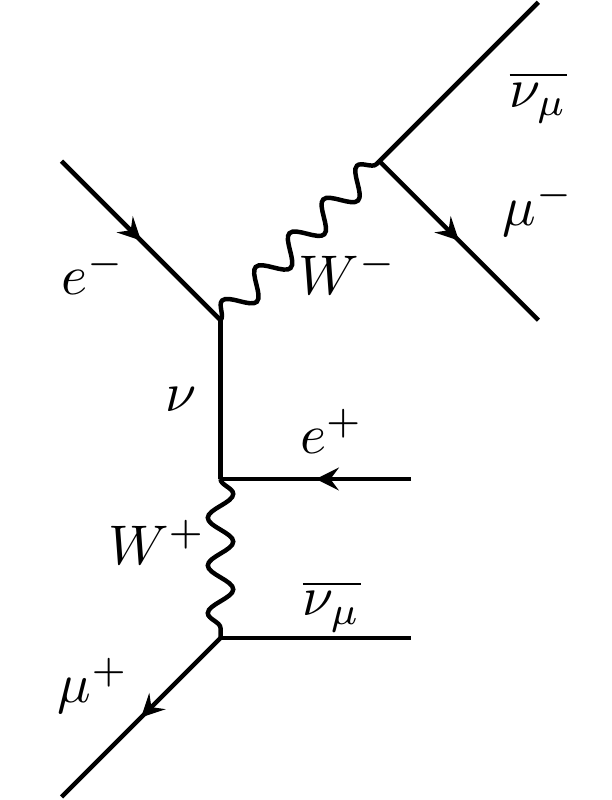} ~~
\includegraphics[width=0.2\textwidth]{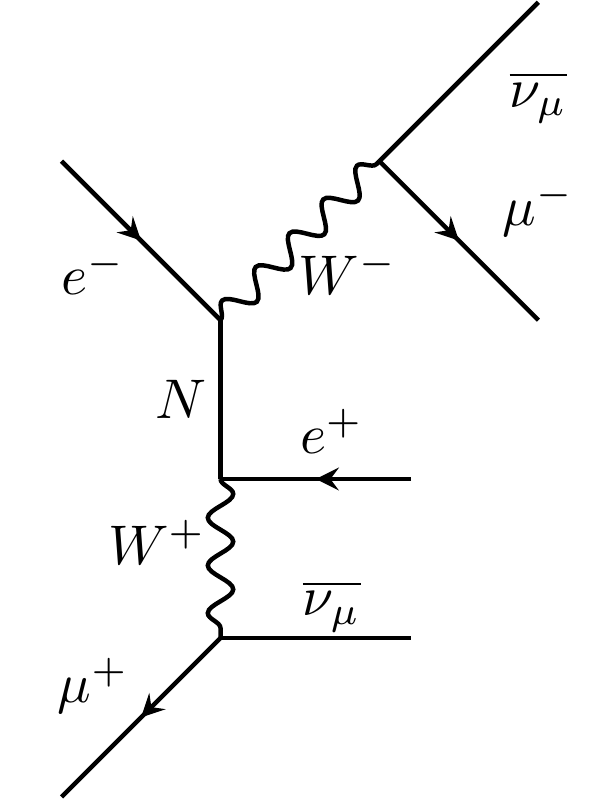} ~~
\includegraphics[width=0.2\textwidth]{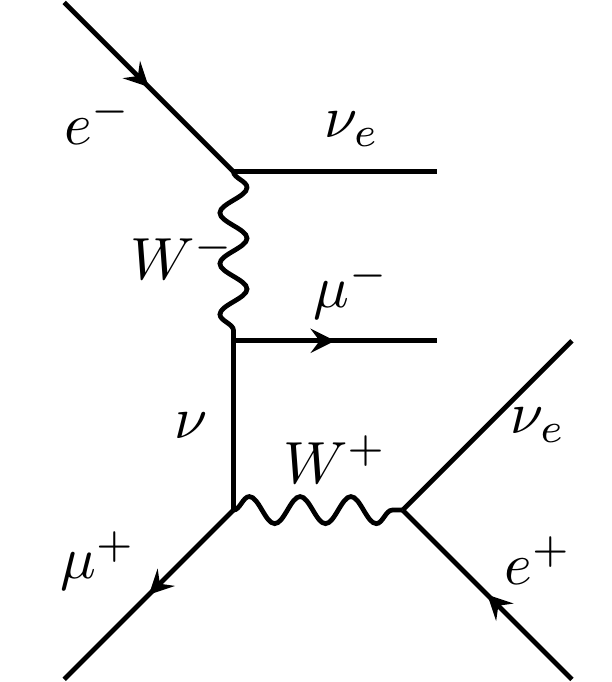} ~~
\includegraphics[width=0.2\textwidth]{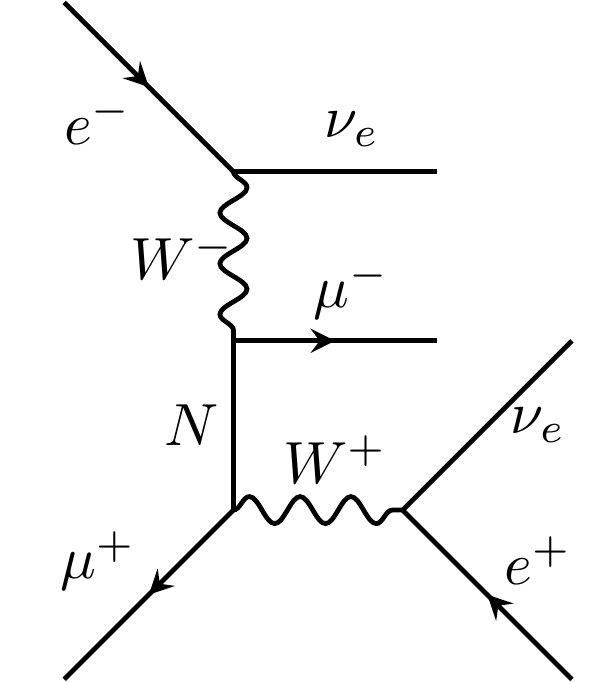}
\end{center}
\caption{Some Feynman diagrams that contribute to the process $e^- \mu^+ \rightarrow \mu^- e^+ + 2~\nu$
{through on- or off-shell $W$.}
The processes shown in the upper panel violate lepton flavor but conserve lepton number whereas the ones in the lower panel violate both lepton flavor as well as lepton number.}
\label{fig2to4}
\end{figure}

{ Before discussing the scattering $e^-\mu^+\to W^+ W^-$, we should make a few more remarks on the case of off-shell $W^+$ and $ W^-$. As already mentioned before, when the CM energy $\sqrt{s}$ is below $2M_W$, the $W^+ W^-$ are only produced off-shell and thus
convert into leptons or hadrons. For instance, if we consider the decay of the $W$ bosons into leptons, we have many possibilities, some of which are shown in Fig.~\ref{fig2to4}. Here we have considered only those processes in which the final state is $e^+ \mu^- ~+~2~\nu$. All of these diagrams violate lepton flavor. The diagrams in the upper row conserve lepton number (LNC), whereas those in the lower row violate lepton number (LNV). However, these two types of processes are experimentally indistinguishable because the two final state neutrinos will appear only as missing energy. Thus, a full analysis of $e^- \mu^+ \rightarrow e^+ \mu^- +$ missing energy will require the 
inclusion of all LFV and LNV topologies, including interference between diagrams with light and heavy neutrino exchange. These amplitudes are suppressed by the mixing between light and heavy neutrinos, as well as by the four-body phase space. An estimate of the scattering cross section yields a value of around $\sigma \approx 10^{-14} - 10^{-15}$ fb. 
In fact, these processes will be a background for the cleaner one-loop process with no missing energy discussed in the previous section. }

\begin{figure}[b]
\begin{center}
\includegraphics[width=0.15\textwidth]{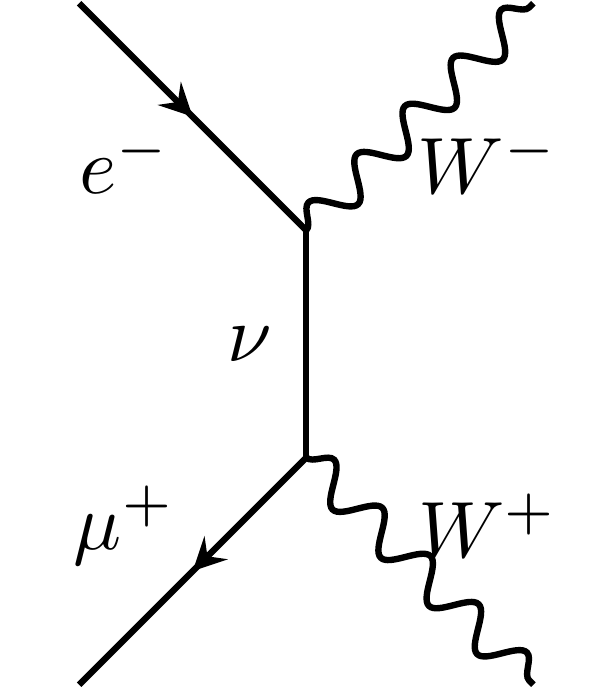} ~~
\includegraphics[width=0.15\textwidth]{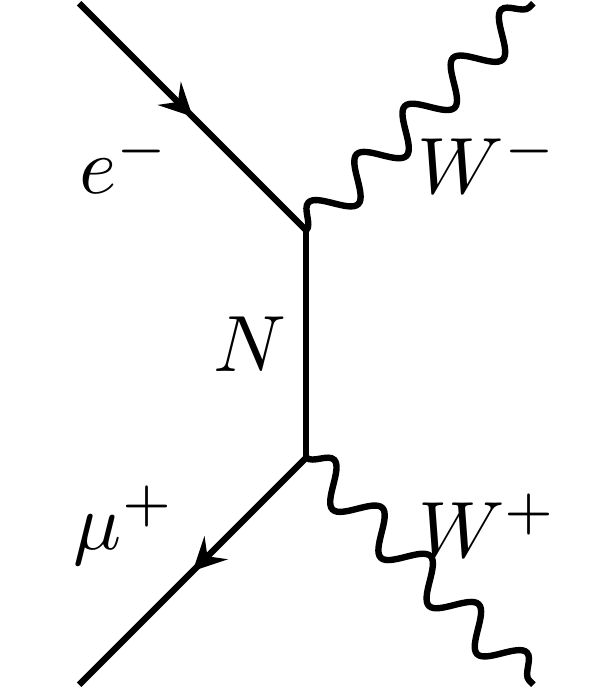}
\end{center}
\caption{Feynman diagrams for the scattering process $e^- (p_1) \mu^+ (p_2) \rightarrow W^- (q_1) W^+ (q_2)$.}
\label{fig:FeynemuWW}
\end{figure}

Now, coming back to the case of $e^-\mu^+$ scattering to produce the on-shell $W^+ W^-$ pair, the corresponding Feynman diagrams are shown in Fig.~\ref{fig:FeynemuWW}. The expression for the squared amplitude, averaged over the spins of the initial particles and summed over the spins of the final particles in the CM frame, is given by
\begin{equation} \left|\mathcal{M}\right|^2 =
\frac{g^{4}}{16 M_W^{4}}
\left(
- 4 M_W^{8}
+ 8 M_W^{6} t
- 4 M_W^{4} s t
- 5 M_W^{4} t^{2}
+ 4 M_W^{2} s t^{2}
+ 2 M_W^{2} t^{3}
- s t^{3}
- t^{4}
\right)
\left|
\sum_{a=1}^{3+n_H}
\frac{U_{e a}^{*} \, U_{\mu a}}{t - M_a^{2}}
\right|^{2}.
\label{Msq}
\end{equation}

As in the previous section, here $n_H$ denotes the total number of heavy sterile neutrinos,
$M_a$ are the neutrino masses, and $U$ is the $n \times n$ unitary neutral lepton mixing matrix, whose $3 \times 3$ sub-block corresponds to the 
PMNS mixing matrix. In the above expression, we neglected the charged lepton masses $m_e, m_\mu$. The Mandelstam variables $s$ and $t$ are defined as $s = (p_1+p_2)^2 = (q_1 + q_2)^2$ and $t=(p_1-q_1)^2 = (p_2 - q_2)^2$, where $p_1, p_2, q_1,$ and $q_2$ are the four-momenta of $e^-$, $\mu^+$, $W^-$, and $W^+$, respectively.

From Eq.~(\ref{Msq}), the total cross section $\sigma(s)$ can be obtained as,
\begin{equation}
\sigma(s) = \frac{1}{16 \pi s}\int_{t_-}^{t_+} \left|\mathcal{M}\right|^2  dt,
\end{equation}
where
\begin{align}
t_\pm &= M_W^2 - \frac{s}{2}(1 \mp \beta), \\
\beta &= \sqrt{1 - \frac{4M_W^2}{s}} .
\end{align}

\subsection{{The case of SM active neutrinos only}}

If only the three SM active neutrinos contribute in Eq.~(\ref{Msq}), then we should recall that their masses $M_a$ $(a=1,2,3)$ are sub-eV and therefore negligible for this high energy process. In this case the amplitude vanishes due to the unitarity of the PMNS matrix $U$: 
\begin{equation}
\sum_{a=1}^3 \frac{U_{e a}^* U_{\mu a}}{t-M_a^2}
\xrightarrow{\, M_a^2 = 0   \,}
 \frac{1}{t}\sum_{a=1}^3 U_{e a}^*U_{\mu a}
= 0, \label{scat:SM}
\end{equation}
where we have used $U^\dag U = I$. Within the SM $\sigma(e^- \mu^+ \rightarrow W^- W^+) =0 $ so that any observation of this process will be an immediate confirmation of new physics and charged lepton flavor violation.

\subsection{The case of SM plus one heavy sterile neutrino}

\begin{figure}[t]
\begin{center}
\includegraphics[width=0.48\textwidth]{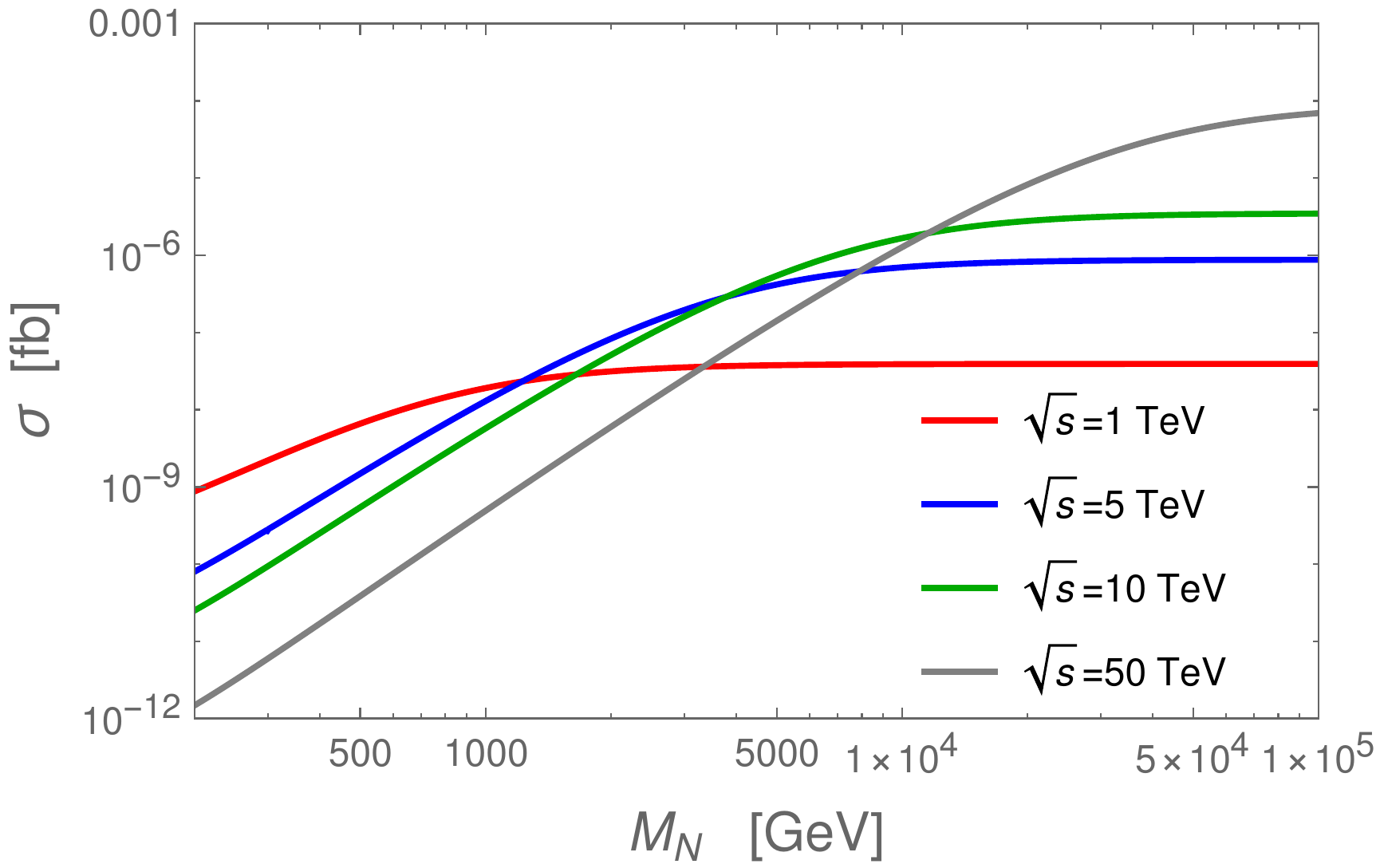} ~~
\includegraphics[width=0.48\textwidth]{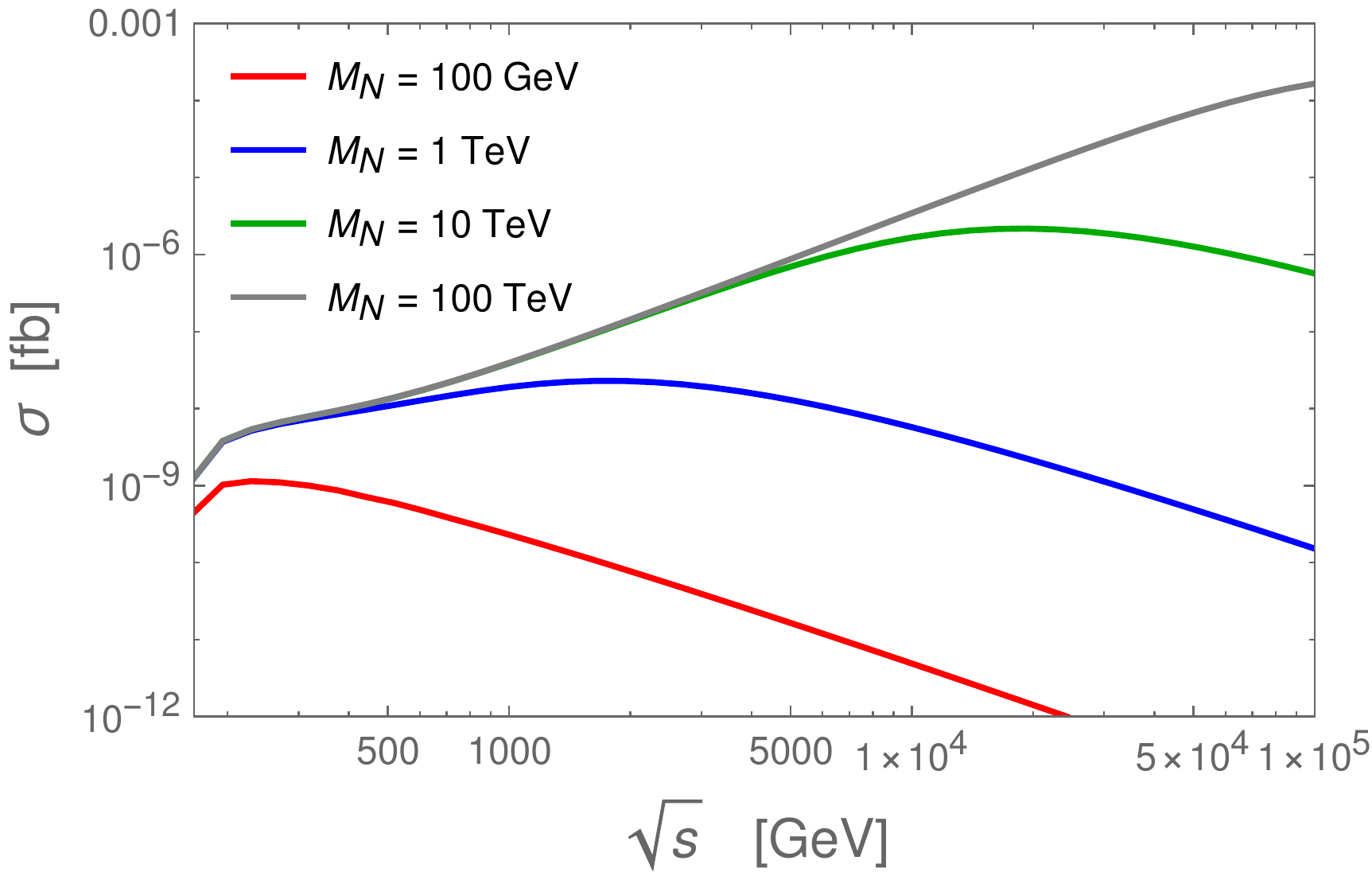}
\end{center}
\caption{Variation of $\sigma(e^- \mu+ \rightarrow W^- W^+)$ with $M_N$ and $\sqrt{s}$ when there is just one heavy sterile neutrino. }
\label{fig:singlesterile}
\end{figure}

Once we have extra heavy neutral leptons that mix with the active light neutrinos, the approximation used in 
Eq.~(\ref{scat:SM}) is no longer valid. We must consider the contributions from the light and heavy neutrinos properly. The neutrino propagator term in Eq.~(\ref{Msq}) becomes
\begin{equation}
\sum_{a=1}^{4}
\frac{U_{e a}^{*} \, U_{\mu a}}{t - M_a^{2}}  
= 
\frac{1}{t} 
\sum_{a=1}^{3} U_{e a}^{*} \, U_{\mu a} 
+ 
\frac{U_{e N}^{*} \, U_{\mu N}}{t - M_N^{2}},
\end{equation}
where the 4th neutrino is simply called $N$, with a large mass $M_N$.
Unitarity implies that $\sum_{a=1}^{3} U_{e a}^{*} \, U_{\mu a}  = - U_{e N}^{*} \, U_{\mu N} $, therefore:
\begin{equation}
\sum_{a=1}^{4}
\frac{U_{e a}^{*} \, U_{\mu a}}{t - M_a^{2}}  
=
\left(
- \frac{1}{t}  + 
\frac{1}{t - M_N^{2}}\right) U_{e N}^{*} \, U_{\mu N}.
\end{equation}
As before, we fix
$|U_{e N}^{*} \, U_{\mu N}| = 10^{-5}$ and study the variation of cross section for the process $e^- \mu+ \rightarrow W^- W^+$ for different values of the mass of the sterile neutrino $M_N$. Our results are shown in Fig.~\ref{fig:singlesterile}.

On the left graph of Fig.~\ref{fig:singlesterile}, we show the dependence of $\sigma$ on $M_N$, for different values of $\sqrt{s}$. The red, blue, green, and gray lines correspond to $\sqrt{s} = 1$ TeV, $5$ TeV, $10$ TeV, and $50$ TeV, respectively. {From this figure we see that, for a given $\sqrt{s}$, the cross section increases with $M_N$ until $M_N \approx \sqrt{s}$, and then saturates for $M_N$ above $\sqrt{s}$.  Another feature is that, for larger $\sqrt{s}$, the cross section reaches lower values at low $M_N$ and larger values at large $M_N$.}

On the right graph of Fig.~\ref{fig:singlesterile}, we show the dependence of $\sigma$ with $\sqrt{s}$, for different values of $M_N$. The red, blue, green, and gray curves correspond to $M_N =$ $100$ GeV, $1$ TeV, $10$ TeV, and $100$ TeV, respectively. From the figure, we see that the cross section increases  with $\sqrt{s}$ up to the point where $\sqrt{s}\sim M_N$, and from then on it decreases with $\sqrt{s}$; the latter is the expected feature for a cross section that respects perturbative unitarity at high energies.

\subsection{Minimal type-I seesaw : SM + 2 heavy sterile neutrinos}

One of the most popular mechanisms to explain the light neutrino masses is the type-I seesaw mechanism, where the SM particle content is extended by heavy Majorana neutrinos $N_R$~\cite{Minkowski:1977sc, Gell-Mann:1979vob, Yanagida:1979as, Mohapatra:1979ia}. We consider the minimal type-I scenario, where we add two Majorana neutrinos $(n_H=2)$ to the SM. 

The heavy Majorana neutrinos couple to the SM neutrinos and the Higgs via Yukawa couplings $Y_\nu$ and generate the dimension-5 Weinberg operators at low energy scales. Once electroweak symmetry is broken, these operators give rise to tiny Majorana masses for the active light neutrinos,
\begin{equation}
M_\nu = -m_D D_N^{-1} m_D^T
\end{equation}

\begin{figure}[t]
\begin{center}
\includegraphics[width=0.48\textwidth]{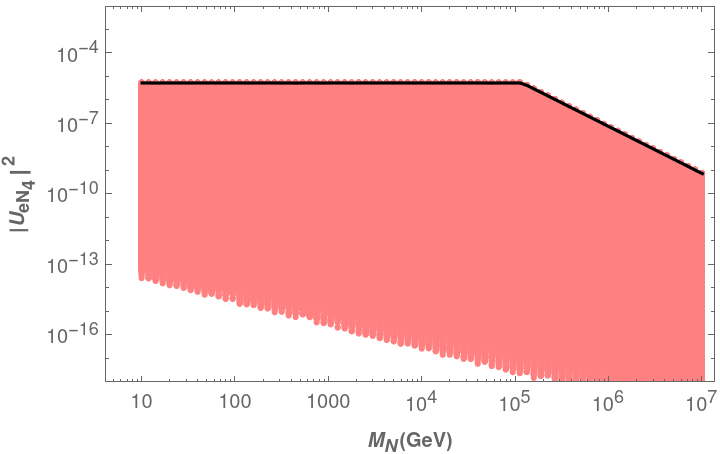} ~~
\includegraphics[width=0.48\textwidth]{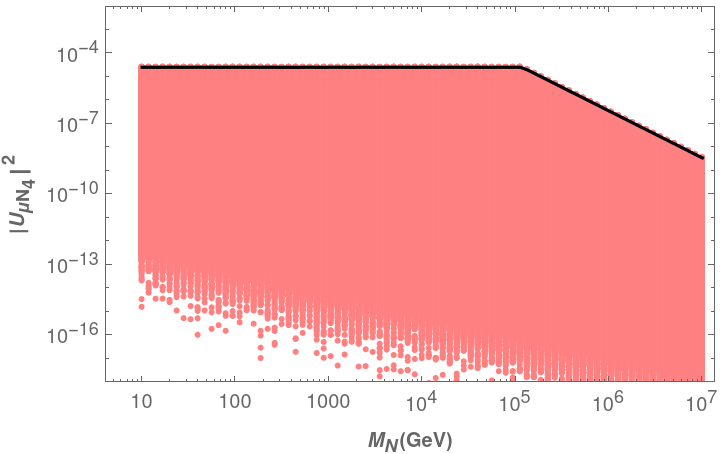}
\end{center}
\caption{Predictions for light–heavy mixing as a function of the heavy neutrino mass in the minimal type-I seesaw model with degenerate heavy Majorana neutrinos, $M_{N_1}=M_{N_2}=M_N$. The region above the black line is disfavored by the bounds on the non-unitarity of the PMNS matrix.}
\label{fig:nonunitary}
\end{figure}

when $m_D \equiv  Y_\nu v/\sqrt{2}$ is much smaller than $D_N$. Here, $v$ is the vacuum expectation value of the SM Higgs, and $D_N = \mathrm{diag}(M_{N_1},M_{N_2})$, where $M_{N_{1,2}}$ are the masses of the heavy Majorana neutrinos. In this case, with $n_H=2$, the lightest active neutrino is massless. To ensure that we use neutrino mixing matrices and masses consistent with oscillation data, we adopt the Casas–Ibarra parametrization for the Dirac neutrino mass matrix~\cite{Casas:2001sr},
\begin{equation}
m_D =  U_\nu \sqrt{D_\nu} R \sqrt{D_N},
\end{equation}
where $U_\nu$ is the $3\times3$ PMNS mixing matrix, $R$ is a $3 \times 2$ complex orthogonal matrix satisfying $R^T R = \mathbb{I}$, and $D_\nu = \mathrm{diag}(M_1,M_2,M_3)$ is the diagonal matrix of light neutrino masses. Such a parametrization also helps us find the maximum light–heavy mixing for a given Majorana neutrino mass, as allowed by the bounds on the non-unitarity of the PMNS matrix~\cite{Blennow:2023mqx}. By choosing the complex angle that parametrizes the orthogonal matrix $R$, we can obtain arbitrarily large values of light–heavy mixing matrix,
\be
U_{\alpha N_{4,5}} = m_D D_N^{-1},
\ee
while simultaneously satisfying the experimental constraints. These limits on $|U_{e N_4}|$ and $|U_{\mu N_4}|$ are shown in Fig.~\ref{fig:nonunitary}. In these plots, we assumed the heavy Majorana neutrinos to be degenerate and also assumed normal hierarchy for the active light neutrinos. Thus we take
\begin{equation}
R = \begin{pmatrix}  
0 & \textrm{cos}(z) & \textrm{sin}(z) \\
0 & -\textrm{sin}(z) & \textrm{cos}(z)
\end{pmatrix} ~~; ~~~~~ D_\nu = \mathrm{diag} \Big(0 , ~~\sqrt{ \Delta m_{\textrm{sol}}^2 }, ~~\sqrt{ \Delta m_{\textrm{sol}}^2 + \Delta m_{\textrm{atm}}^2 }  \Big).
\end{equation}
We have also fixed the neutrino mass squared differences, mixing angles and Dirac CP phase to be the best fit values as indicated by the global analysis of the oscillation data~\cite{Esteban:2024eli} and the Majorana CP phase is taken to be $0$. The regions above the pink points in Fig.~\ref{fig:nonunitary} are disfavored by the bounds on the non-unitarity of the PMNS matrix.

\begin{figure}[h]
\begin{center}
\includegraphics[width=0.5\textwidth]{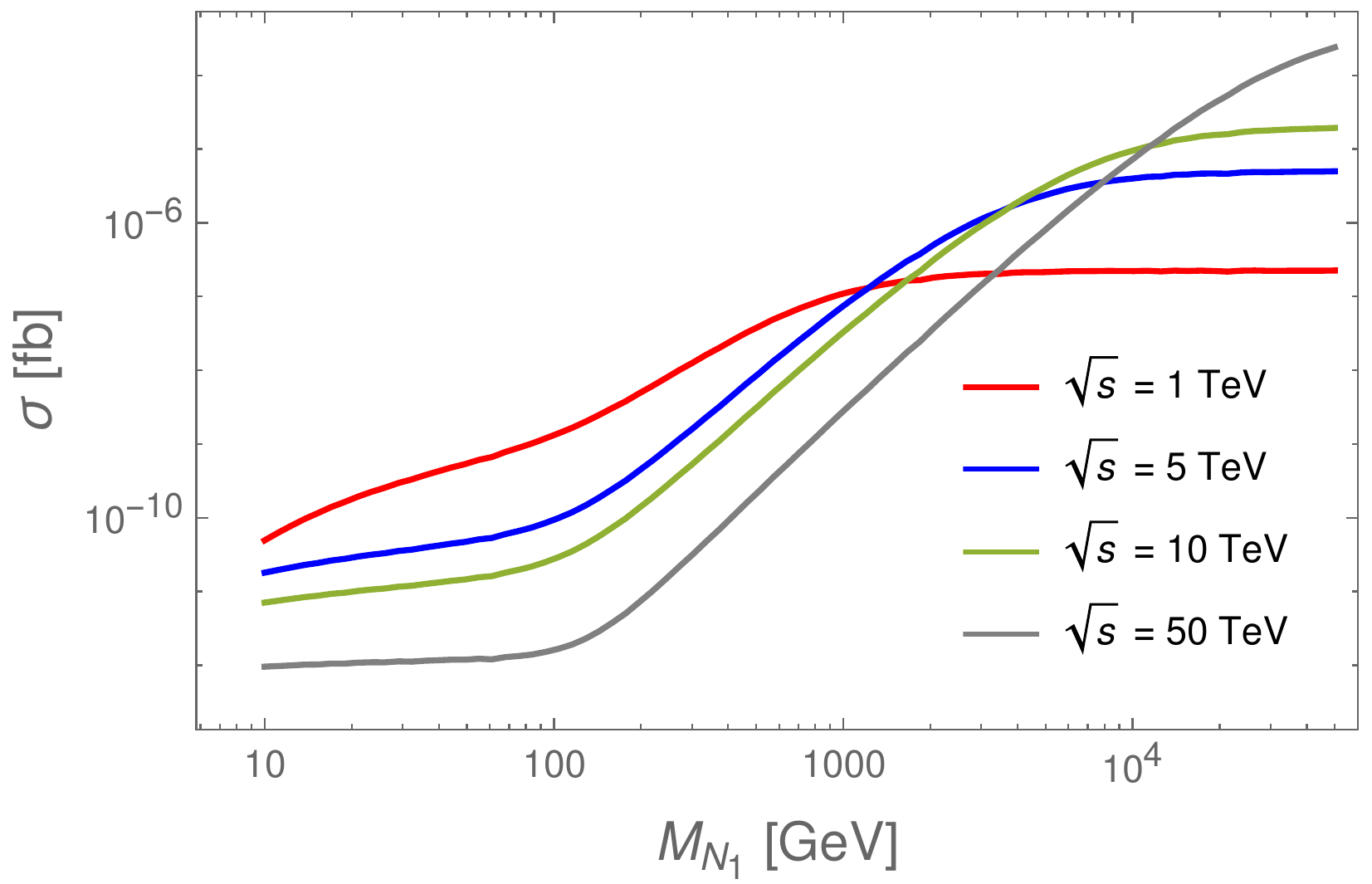} 
\end{center}
\caption{Variation of $\sigma$ with $M_{N_1}=M_{N_2}=M_N$ for different values of $\sqrt{s}$ in the minimal type-I seesaw mechanism.}
\label{fig:sigma_sqrts}
\end{figure}

\begin{figure}[h]
\begin{center}
\includegraphics[width=0.5\textwidth]{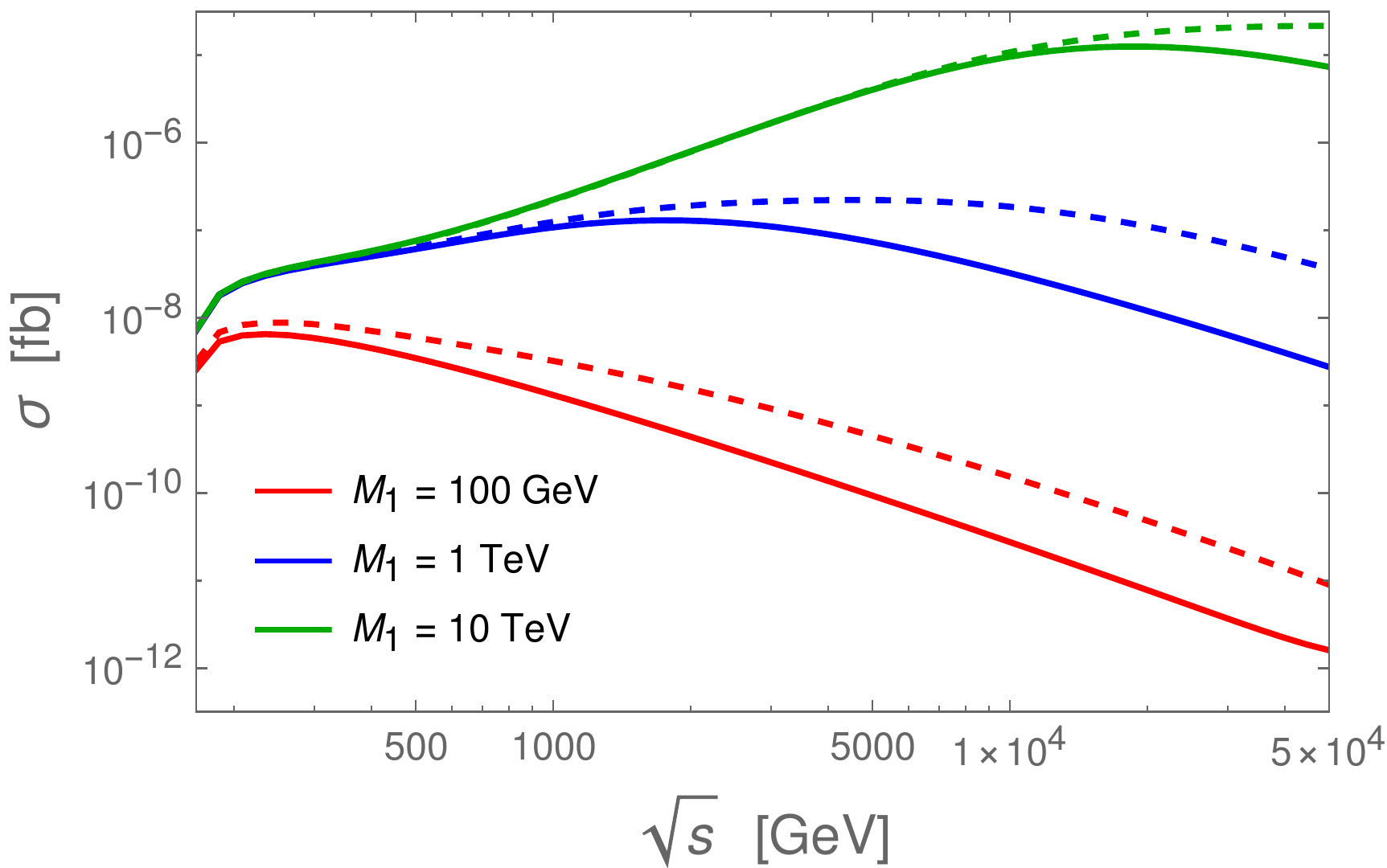} 
\end{center}
\caption{Variation of $\sigma$ with $\sqrt{s}$ for different values of $M_{N_1}$. The solid lines correspond to the case where the two heavy neutrinos are degenerate whereas the dashed line correspond to the case where the mass of the second heavy neutrino, $M_{N_2} = 5~ M_{N_1}$.}
\label{fig:sigma_MN}
\end{figure}

In Fig.~\ref{fig:sigma_sqrts}, we show the behavior of $\sigma (e^- \mu^+ \rightarrow W^- W^+)$ with $M_{N_1}$ for different values of $\sqrt{s}$, taking the maximal mixing allowed by the bounds shown in Fig.~\ref{fig:nonunitary}. Similarly, in Fig.~\ref{fig:sigma_MN}, we show the behavior of $\sigma$ with $\sqrt{s}$, for different values of $M_{N_1}$. The solid lines correspond to the case where the two heavy neutrinos are degenerate, whereas the dashed line corresponds to the case where the mass of the second heavy neutrino, $M_{N_2} = 5~ M_{N_1}$.

 \section{Summary and Conclusions}

We studied high energy $e^-\mu^+$ collisions that can provide signals of charged lepton flavor violation (cLFV), specifically due to the existence of heavy neutral leptons that mix with the Standard neutrinos.
We considered alternatively the simple representative case of one heavy sterile neutrino (in Secs.~\ref{sec:Box} and III) and of two heavy sterile neutrinos, common to the minimal type-I seesaw scenario (in Sec.~III).

In Sec.~\ref{sec:Box} we calculated the purely leptonic cLFV process $e^- \mu^+ \to e^+ \mu^-$ for CM energies  $m_e \ll \sqrt{s} \leq 2 M_W$. 
In the scenario of extra heavy neutral leptons, this process occurs at one loop level of $box$ diagrams involving all neutral leptons. We numerically evaluated these box diagrams in the mentioned energy range, which involve the high-energy extension of the Inami-Lim functions. It turns out that, due to the unitarity of the $4\times 4$ or $5 \times 5$ neutrino mixing matrix, the box diagram amplitudes are quartic in the heavy-light mixing elements $U_{\ell N}$, which leads to a significant suppression, cf.~Fig.~\ref{FigsigBox}. Indeed, we obtain cross sections for $\sigma(e^-\mu^+\to e^+\mu-)$ below $10^{-16}$ fb, while a Standard cross section such as $\sigma(e^-\mu^+\to \nu_e\bar\nu_\mu)$ at 100 GeV is $\sim 10^5$ fb.
In Sec.~III, we studied the scattering process $e^- \mu^+$ for CM energies $\sqrt{s} > 2 M_W$, where the channel $e^- \mu^+ \to W^+ W^-$ becomes kinematically accessible, leading to the production of on-shell $W$ bosons. We determined the cross sections both in the case of the SM extended by a single sterile neutrino (cf.~Fig.~\ref{fig:singlesterile}) as well as in the context of the minimal type-I seesaw model {with two extra neutral leptons} (cf.~Figs.~\ref{fig:sigma_sqrts} and \ref{fig:sigma_MN}). In the minimal seesaw framework, we obtained the maximum achievable cross sections by using the largest allowed values of the light–heavy neutrino mixing (cf.~Fig.~\ref{fig:nonunitary}), consistent with current experimental constraints on the non-unitarity of the PMNS matrix. Any observation of $W^+ W^-$ production from $e^- \mu^+$ scattering would be an immediate indication of new physics and could potentially hint at the existence of {heavy sterile neutrino(s).} Thus, a future $e^- \mu^+$ collider could test charged lepton flavor violation and the sterile neutrino hypothesis simultaneously.

\begin{acknowledgments}
\noindent

{This work was supported in part by Fondecyt (Chile) grants 1250776 and 1220095, and by ANID CCTVal CIA250027. C.S.K. also received support  by Basic Science Research Program through the National Research Foundation of Korea (NRF)
funded by the Ministry of Education (RS-2022-NR070836, RS-2026-25471898).}

\end{acknowledgments}


\appendix

\section{Additional formulas for $e^- \mu^+ \to \mu^- e^+$ box diagrams}
\label{app:Box}

Similar to the expression for $\cF^{(1,WW)}$  for the box diagram of type $t=1$ given in Eq.~\ref{F1a2}, the diagrams of the types $t=2,3,4$ are then
\bes
\label{Fja}
\bea
\cF^{(2,WW)} &=& 
\frac{i}{16 \pi^2} \frac{g^4}{2M_W^2} \nonumber
\\ 
&&\left\{
(-1) \bF^{(2)}_{1} \left[ \bu(\ell') \gamma^{\mu} P_L u(k) \right] \left[ \bu(k') \gamma_{\mu} P_R u(\ell) \right] - 2 \bF^{(2)}_2(1-x,1-y) \frac{1}{M_W^2} \left[\bu(\ell') \lslash P_L u(k) \right] \left[ \bu(k') \kslash P_R u(\ell) \right] \right\}
\nonumber\\
\label{F2a} \\
\cF^{(3,WW)} &=& \frac{i}{16 \pi^2} \frac{g^4 }{2 M_W^2}
\sqrt{x_a x_b}\ \ 
\bF^{(3)}_{2}(1,1) \left[ \bu(\ell') \gamma^{\mu} P_L u(k) \right] \left[ \bu(k') \gamma_{\mu} P_R u(\ell) \right],
\label{F3a} \\
\cF^{(4,WW)} &=& \frac{i}{16 \pi^2} \frac{g^4 }{2 M_W^2}
\sqrt{x_a x_b}\ \  
\bF^{(4)}_{2}(1,1) \left[ \bu(\ell') \gamma^{\mu} P_L u(k) \right] \left[ \bu(k') \gamma_{\mu} P_R u(\ell) \right].
\label{F4a}
\eea \ees
Here, the integrals $\bF^{(t)}_1$ and $\bF^{(t)}_2$, appearing in the type \((t,WW)\) diagrams, are defined as in Eqs.~(\ref{bF1def})
 \be
\bF^{(t)}_1 = \int dx dy dz \frac{1}{\Delta^{(t)}_W}, \quad
\bF^{(t)}_2(f_1,f_2) =    \int dx dy dz \frac{f_1 f_2}{\left( \Delta^{(t)}_W \right)^2},
\label{bFjdef} \ee
where again the integration is over $0 \leq x + y + z \leq 1$ and $0 \leq x,y,z \leq 1$, and the expressions $\Delta^{(t)}_W$ in the denominators, for diagrams of type $t=2,3,4$, are
\bes
\label{DeltaWj}
\bea
\Delta^{(2)}_W &=& \left[ (1-x-y) (1 - u_W z) + u_W z^2 + x y (- s_W) + x x_a + y x_b \right] - i \varepsilon_W - i \varepsilon_N.
\label{DeltaW2} \\
\Delta^{(3)}_W &=& \left[ (1-x-y) (1 - t_W z) + t_W z^2 + x y (- s_W) + x x_a + y x_b \right] - i \varepsilon_W - i \varepsilon_N.
\label{DeltaW3} \\
\Delta^{(4)}_W &=& \left[ (1-x-y) (1 - t_W z) + t_W z^2 + x y (- u_W) + x x_a + y x_b \right] - i \varepsilon_W - i \varepsilon_N.
\label{DeltaW4} \eea \ees

Also, as mentioned before, the diagrams of type 
($t$,{\small GW}), ($t$,{\small WG}) 
and ($t$,{\small GG}) are obtained from those of type
($t$,{\small WW}) by replacing the $W$-propagator by the corresponding Goldstone propagator. 
In this way, for the boxes of type 1 with Goldstones we have
\bes
\label{F1bcd1}
\bea
\lefteqn{
  \cF^{(1,{GW})} =  } 
\nonumber\\ &&
\frac{g^4}{4}\int \frac{d^4 p}{(2 \pi)^4} \frac{(-1)^2 \left[ \bu(\ell') M_a P_R (\pslash +\lpslash+M_a) \gamma^{\beta} P_L v(k') \right] \left[ \bv(\ell) \gamma_{\beta} P_L (\pslash + \kslash + M_b) M_b P_L u(k) \right]}{ \left((p+\ell')^2-M_a^2 + i \Gamma_a M_a \right) \left( (p+k)^2 - M_b^2  + i \Gamma_b M_b \right) \left( (p+k+\ell)^2 - M_W^2 + i \Gamma_W M_W\right) \left( p^2 - M_W^2 + i \Gamma_W M_W \right) },
\nonumber\\
\label{F1b1} \\
\lefteqn{\cF^{(1,{WG})} =}
\nonumber\\ &&
\frac{g^4}{4}\int \frac{d^4 p}{(2 \pi)^4} \frac{(-1)^2 \left[ \bu(\ell') \gamma^{\alpha} P_L (\pslash +\lpslash+M_a) M_a P_L v(k') \right] \left[ \bv(\ell) M_b P_R (\pslash + \kslash + M_b) \gamma_{\alpha} P_L u(k) \right]}{ \left((p+\ell')^2-M_a^2 + i \Gamma_a M_a \right) \left( (p+k)^2 - M_b^2  + i \Gamma_b M_b \right) \left( (p+k+\ell)^2 - M_W^2 + i \Gamma_W M_W\right) \left( p^2 - M_W^2 + i \Gamma_W M_W \right) },
\nonumber\\
\label{F1c1} \\
\lefteqn{\cF^{(1,GG)} =}
\nonumber\\ &&
\frac{g^4}{4}\int \frac{d^4 p}{(2 \pi)^4} \frac{(-1)^3 \left[ \bu(\ell') M_a P_R (\pslash +\lpslash+M_a) M_a P_L v(k') \right] \left[ \bv(\ell) M_b P_R (\pslash + \kslash + M_b) M_b P_L u(k) \right]}{ \left((p+\ell')^2-M_a^2 + i \Gamma_a M_a \right) \left( (p+k)^2 - M_b^2 + i \Gamma_b M_b \right) \left( (p+k+\ell)^2 - M_W^2 + i \Gamma_W M_W \right) \left( p^2 - M_W^2 + i \Gamma_W M_W \right) }.
\nonumber\\
\label{F1d1}
\eea \ees
The corresponding integrals for the box types
$t=2,3,4$ with Goldstones are obtained in analogy to those in Eqs.~(\ref{Fja}).

Thus for the boxes ($t$,{\small GW}) the integrals are:

\bes
\label{Fjb}
\bea
\cF^{(1,GW)} &=& \frac{i}{16 \pi^2} \frac{ g^4 }{4 M_W^2} x_a x_b (-1) \bF^{(1)}_2(1,1) \left[ \bu(\ell') \gamma^{\mu} P_L u(k) \right] \left[ \bu(k') \gamma_{\mu} P_R u(\ell) \right],
\label{F1b2} 
\\
\cF^{(2,GW)} &=& \frac{i}{16 \pi^2} \frac{g^4 }{4 M_W^2} x_a x_b (-1) \bF^{(2)}_2(1,1) \left[ \bu(\ell') \gamma^{\mu} P_L u(k) \right] \left[ \bu(k') \gamma_{\mu} P_R u(\ell) \right],
\label{F2b2} 
\\
\cF^{(3,GW)} &=& \frac{i}{16 \pi^2} \frac{g^4}{4 M_W^2} \sqrt{x_a x_b} {\Bigg \{}
\left( \bF^{(3)}_1 + s_W \bF^{(3)}_2(z,z) \right)
\left[ \bu(\ell') \gamma^{\mu} P_L u(k) \right] \left[ \bu(k') \gamma_{\mu} P_R u(\ell) \right]
\nonumber
\\ &&
+ \bF^{(3)}_2(x,y) \frac{1}{M_W^2} \left[\bu(\ell') \gamma^{\mu} \kslash P_R u(\ell) \right] \left[ \bu(k') \gamma_{\mu} \lslash P_L u(k) \right]
\nonumber\\ &&
+2 \bF^{(3)}_2(y,z)  \frac{1}{M_W^2} \left[\bu(\ell') \kpslash \kslash P_R u(\ell) \right] \left[ \bu(k') P_L u(k) \right]
\nonumber\\ &&
+2 \bF^{(3)}_2(z,x)  \frac{1}{M_W^2} \left[\bu(\ell')  P_R u(\ell) \right] \left[ \bu(k') \lpslash \lslash P_L u(k) \right]
    {\Bigg \}},
\label{F3b2} 
\\
\cF^{(4,GW)} &=& \frac{i}{16 \pi^2} \frac{g^4 }{4 M_W^2}
\sqrt{x_a x_b}{\Bigg \{}
\left( \bF^{(4)}_1 + u_W \bF^{(4)}_2(z,z) \right)
\left[ \bu(\ell') \gamma^{\mu} P_L u(k) \right] \left[ \bu(k') \gamma_{\mu} P_R u(\ell) \right]
\nonumber\\ &&
- \bF^{(4)}_2(x,y) \frac{1}{M_W^2} \left[\bu(\ell') \gamma^{\mu} \kpslash P_R u(\ell) \right] \left[ \bu(k') \lslash \gamma_{\mu} P_L u(k) \right]
\nonumber\\ &&
+2 \bF^{(4)}_2(y,z)  \frac{1}{M_W^2} \left[\bu(\ell') \kslash \kpslash P_R u(\ell) \right] \left[ \bu(k') P_L u(k) \right]
\nonumber\\ &&
+2 \bF^{(4)}_2(z,x)  \frac{1}{M_W^2} \left[\bu(\ell')  P_R u(\ell) \right] \left[ \bu(k') \lslash \lpslash P_L u(k) \right]
    {\Bigg \}}.
\label{F4b2} \eea \ees

Similarly, for the boxes ($t$,\small{WG}) 
the integrals  are:
\bes
\label{Fjc}
\bea
\cF^{(1,WG)} &=& \cF^{(1,GW)}, \; \cF^{(2,WG)} = \cF^{(2,GW)},
\label{F12c2} 
\\
\cF^{(3,WG)} &=&  \frac{i}{16 \pi^2} \frac{g^4 }{4 M_W^2} \sqrt{x_a x_b} {\Bigg \{}
\left( \bF^{(3)}_1 + s_W \bF^{(3)}_2(1-x-z,1-y-z) \right)
\left[ \bu(\ell') \gamma^{\mu} P_L u(k) \right] \left[ \bu(k') \gamma_{\mu} P_R u(\ell) \right]
\nonumber\\ &&
+ \bF^{(3)}_2(x,y) \frac{1}{M_W^2} \left[\bu(\ell') \kslash \gamma^{\mu} P_R u(\ell) \right] \left[ \bu(k') \lslash \gamma_{\mu} P_L u(k) \right] {\Bigg \}},
\label{F3c2} 
\\
\cF^{(4,WG)} &=&  \frac{i}{16 \pi^2} \frac{g^4}{4M_W^2} \sqrt{x_a x_b} {\Bigg \{}
\left( \bF^{(4)}_1 + u_W \bF^{(4)}_2(1-x-z,1-y-z) \right)
\left[ \bu(\ell') \gamma^{\mu} P_L u(k) \right] \left[ \bu(k') \gamma_{\mu} P_R u(\ell) \right]
\nonumber\\ &&
- \bF^{(4)}_2(x,y) \frac{1}{M_W^2} \left[\bu(\ell') \kpslash \gamma^{\mu} P_R u(\ell) \right] \left[ \bu(k') \gamma_{\mu} \lslash P_L u(k) \right] {\Bigg \}}.
\label{F4c2}
\eea \ees
And finally for the boxes ($t$,GG) 
the integrals  are:
\bes
\label{Fjd}
\bea
\cF^{(1,GG)} &=& \frac{i}{16 \pi^2} \frac{g^4}{4 M_W^2}x_a x_b
\left\{ -\frac{1}{2}  \bF^{(1)}_1 \left[ \bu(\ell') \gamma^{\mu} P_L u(k) \right] \left[ \bu(k') \gamma_{\mu} P_R u(\ell) \right] - \bF^{(1)}_2(x,y) \frac{1}{M_W^2} \left[\bu(\ell') \kslash P_L v(k') \right] \left[ \bv(\ell) \lpslash P_L u(k) \right]
\right\},
\nonumber\\
\label{F1d2} \\
  \cF^{(2,GG} &=& \frac{i}{16 \pi^2} \frac{g^4}{4 M_W^2} x_a x_b
\left\{ -\frac{1}{2}  \bF^{(2)}_1 \left[ \bu(\ell') \gamma^{\mu} P_L u(k) \right] \left[ \bu(k') \gamma_{\mu} P_R u(\ell) \right] - \bF^{(2)}_2(x,y) \frac{1}{M_W^2} \left[\bu(\ell') \lslash P_L u(k) \right] \left[ \bu(k') \kslash P_R u(\ell) \right]
\right\},
\nonumber\\
\label{F2d2} \\
\cF^{(3,GG)} &=&  \frac{i}{16 \pi^2} \frac{g^4}{4 M_W^2}(x_a x_b)^{3/2} \frac{1}{2} \bF^{(3)}_2(1,1) \left[ \bu(\ell') \gamma^{\mu} P_L u(k) \right] \left[ \bu(k') \gamma_{\mu} P_R u(\ell) \right],
\label{F3d2} \\
\cF^{(4,GG)} &=& \frac{i}{16 \pi^2} \frac{g^4}{4 M_W^2} (x_a x_b)^{3/2} \frac{1}{2} \bF^{(4)}_2(1,1) \left[ \bu(\ell') \gamma^{\mu} P_L u(k) \right] \left[ \bu(k') \gamma_{\mu} P_R u(\ell) \right].
\label{F4d2}
\eea \ees


The full amplitude ${\cal A}_{\rm box}$ is the sum of the full amplitudes $\cF^{(t,p)}_f$ ($t=1,2,3,4$; $p=$ {\small WW, GW, WG, GG}) , cf.~the expressions in Eqs.~(\ref{F1a2}), (\ref{Fja}), (\ref{Fjb})-(\ref{Fjd}) and (\ref{cFj12})-(\ref{cFj34}). The structure of the terms quartic in the spinors is in all but two terms such that it contains only the spinors $u$ (not $v$): $u(k)$, $u(\ell')$, $\bu(\ell')$ and $\bu(k')$. This structure then makes possible a direct evaluation of the traces appearing in $\langle |\sum \cF^{(t,p)}|^2 \rangle$, using the relations (\ref{normu}). These expressions were obtained by means of the following Fierz transformations:
\bes
\label{Fierz1}
\bea
\left[ \bu(\ell') \gamma^{\mu} P_L v(k') \right] 
\left[ \bv(\ell) \gamma_{\mu} P_L u(k) \right] & = &
- \left[ \bu(\ell') \gamma^{\mu} P_L u(k) \right] 
\left[ \bv(\ell) \gamma_{\mu} P_L v(k') \right],
\label{Fi1a} \\
\left[ \bu(\ell') \sigma^{\mu \nu} P_R v(\ell) \right] 
\left[ \bv(k') \sigma_{\mu \nu} P_L u(k) \right] & = & 0,
\label{Fi1b} \\
\left[ \bu(\ell') P_R u(\ell) \right] 
\left[ \bu(k') P_L u(k) \right] & = &
\frac{1}{2} \left[ \bu(\ell') \gamma^{\mu} P_L u(k) \right] 
\left[ \bu(k') \gamma_{\mu} P_R u(\ell) \right].
\label{Fi1c} \eea \ees
In the Fierz relation (\ref{Fi1a}) we also applied the ${\hat C}$-conjugation such that $v={\hat C} \gamma^0 u^{\ast}$:
\be
\left[ \bv(\ell) \gamma_{\mu} P_L v(k') \right] =
\left[ \bu(k') \gamma_{\mu} P_R u(\ell) \right],
\label{Crel} \ee
leading to
\be
\left[ \bu(\ell') \gamma^{\mu} P_L v(k') \right] 
\left[ \bv(\ell) \gamma_{\mu} P_L u(k) \right] = 
- \left[ \bu(\ell') \gamma^{\mu} P_L u(k) \right] 
\left[ \bu(k') \gamma_{\mu} P_R u(\ell) \right].
\label{Fi1aa} \ee
The Fierz relation (\ref{Fi1c}) can be applied optionally, as there is acceptable spinor structure on both sides of that relation.

In addition, we note that in the expressions (\ref{F1a2}) and (\ref{F1d2}) we have a more complicated spinor structure in the term  $\left[\bu(\ell') \kslash P_L v(k') \right]$ $\left[ \bv(\ell) \lpslash P_L u(k) \right]$. In order to transform it into the needed structure (with: $u(k)$, $u(\ell')$, $\bu(\ell')$ and $\bu(k')$), no simple Fierz transformation is available, as no contraction of the type $\gamma^{\mu} \ldots \gamma_{\mu}$ appears there. Nonetheless, by applying the generalized Fierz transformation approach as described in \cite{Itzykson:Zuber}, and using also the ${\hat C}$- transformed Eq.~(\ref{Crel}) for $[ \bv(\ell) \kslash P_L v(k') ]$, the following relation is obtained:
\bea
\left[\bu(\ell') \kslash P_L v(k') \right] \left[ \bv(\ell) \lpslash P_L u(k) \right] & = &
\frac{1}{2} u \left[ \bu(\ell') \gamma^{\mu} P_L u(k) \right] 
\left[ u(k') \gamma_{\mu} P_R u(\ell) \right] + i \epsilon^{\mu \nu \delta \eta} k_{\mu} \ell'_{\nu} \left[ \bu(\ell') \gamma_{\eta} P_L u(k) \right] 
  \left[ u(k') \gamma_{\delta} P_R u(\ell) \right]
\nonumber\\
&&+ \left[\bu(\ell') \lpslash P_L u(k) \right] \left[ \bu(k') \kslash P_R u(\ell) \right]  + \frac{1}{2} \left[\bu(\ell') \kslash P_L u(k) \right] \left[ \bu(k') \lpslash P_R u(\ell) \right],
\label{Fi2} \eea
where the last two terms on the right-hand 
vanish in the limit $m_e, m_\mu\to 0$.
The Mandelstam variables $s,t, $ and $u$ are defined in Eq.~\ref{Mand}.

It can also be checked that the term with $\epsilon^{\mu \nu \delta \eta}$ in $|\cF|^2$ only enter in combinations that vanish.

Furthermore, the following spinor terms appearing in Eqs.~(\ref{F4b2}) (\ref{F3c2})-(\ref{F4c2}) can be reduced to other terms already appearing in all these amplitudes:
\bes
\label{red}
\bea
\left[\bu(\ell') \kslash \kpslash P_R u(\ell) \right] \left[ \bu(k') P_L u(k) \right] & = &
- t \left[ \bu(\ell') P_R u(\ell) \right] 
\left[ \bu(k') P_L u(k) \right] -   \left[\bu(\ell') \kpslash \kslash P_R u(\ell) \right] \left[ \bu(k') P_L u(k) \right],
\label{t8} \\
\left[\bu(\ell')  P_R u(\ell) \right] \left[ \bu(k') \lslash \lpslash P_L u(k) \right]
& = &
- t \left[ \bu(\ell') P_R u(\ell) \right] 
\left[ \bu(k') P_L u(k) \right] -   \left[\bu(\ell') P_R u(\ell) \right] \left[ \bu(k') \lpslash \lslash P_L  u(k) \right],
\label{t9} \\
 \left[\bu(\ell') \kslash \gamma^{\mu} P_R u(\ell) \right] \left[ \bu(k') \lslash \gamma_{\mu} P_L u(k) \right]
 & = &
 - s \left[ \bu(\ell') \gamma^{\mu} P_L u(k) \right] 
 \left[ u(k') \gamma_{\mu} P_R u(\ell) \right]
 + \left[\bu(\ell') \gamma^{\mu} \kslash P_R u(\ell) \right] \left[ \bu(k) \gamma_{\mu} \lslash P_L u(k) \right],
\nonumber\\
 \label{t4} \\
 \left[\bu(\ell') \kpslash \gamma^{\mu} P_R u(\ell) \right] \left[ \bu(k') \gamma_{\mu} \lslash P_L u(k) \right]
 & = & u \left[ \bu(\ell') \gamma^{\mu} P_L u(k) \right] 
 \left[ u(k') \gamma_{\mu} P_R u(\ell) \right]
 +  \left[\bu(\ell') \gamma^{\mu} \kpslash P_R u(\ell) \right] \left[ \bu(k') \lslash \gamma_{\mu} P_L u(k) \right].
 \nonumber\\
 \label{t7} \eea \ees

\bibliographystyle{utphys_jr} 
\bibliography{References}

\end{document}